\documentclass[aps, pra, 10pt, twocolumn, superscriptaddress, longbibliography, floatfix]{revtex4-1}
\usepackage[nointlimits]{amsmath}
\usepackage{amssymb, cancel}
\usepackage{graphicx, nicefrac}
\usepackage{subfigure}
\usepackage{color}
\usepackage{stackengine}
\usepackage{txfonts}
\usepackage{MnSymbol}
\usepackage{mathtools}
\usepackage{hyperref}
\hypersetup{colorlinks=true, citecolor=blue, linkcolor=blue}

\graphicspath{{./figures/}}
\def\bra#1{{\left\langle #1 \right|}}
\def\ket#1{{\left| #1 \right\rangle}}
\newcommand{\ie}{\emph{i.e.}}
\newcommand{\eg}{\emph{e.g.}}

\newcommand{\dg}{{^\dagger}}

\begin{document}
\title{Establishing trust in quantum computations}
\author{Timothy Proctor}
\email{tjproct@sandia.gov}
\author{Stefan Seritan}
\author{Erik Nielsen}
\author{Kenneth Rudinger}
\author{Kevin Young}
\author{Robin Blume-Kohout}
\affiliation{Quantum Performance Laboratory, Sandia National Laboratories, Albuquerque, NM 87185, USA and Livermore, CA 94550, USA}
\author{Mohan Sarovar}
\email{mnsarov@sandia.gov}
\affiliation{Sandia National Laboratories, Livermore, CA 94550}

\begin{abstract}
Quantum computing hardware has grown sufficiently complex that it often can no longer be simulated by classical computers, but its computational power remains limited by errors. These errors corrupt the results of quantum algorithms, and it is no longer always feasible to use classical simulations to directly check the correctness of quantum computations. Without practical methods for quantifying the accuracy with which a quantum algorithm has been executed, it is difficult to establish trust in the results of a quantum computation. Here we solve this problem, by introducing a simple and efficient technique for measuring the fidelity with which an as-built quantum computer can execute an algorithm. Our technique converts the algorithm's quantum circuits into a set of closely related ``mirror circuits'' whose success rates can be efficiently measured. It enables measuring the fidelity of quantum algorithm executions both in the near-term, with algorithms run on hundreds or thousands of physical qubits, and into the future, with algorithms run on logical qubits protected by quantum error correction.
\end{abstract}

\maketitle

Quantum computers with hundreds of qubits are becoming available, and error rates on quantum logic operations are rapidly decreasing \cite{Corcoles2020-vn, Arute2019-mk}. This is making it possible to execute increasingly complex quantum algorithms, and quantum advantage on artificial problems has now been demonstrated \cite{Arute2019-mk, Zhong2020-rk}. Marquee quantum algorithms for factoring and quantum simulation likely require large quantum circuits, which can only be implemented with low error on logical qubits protected by quantum error correction. However, there is speculation that quantum computers may soon outperform state-of-the-art classical computers on some scientifically important problems \cite{Preskill2018-jz, Bharti2022-rz,Nam_2020,Yeter-Aydeniz_2020,Tazhigulov_2022}. This possibility magnifies a long-standing and foundational problem in quantum computing: how can the results of a quantum algorithm executed on an imperfect, noisy quantum computer be verified as correct or accurate?

Establishing trust in the results of quantum computations, \eg, to verify any claims of quantum advantage, is a multi-faceted task. Arguably, the most important and challenging part of this endeavor is establishing that the algorithm's quantum component---its quantum circuits---were implemented with sufficient accuracy, despite the presence of inevitable hardware errors and noise. This requires verifying the correct operation of a complex system that is infeasible to simulate using even the most powerful supercomputers. The noise and errors in a quantum computer's individual components, \ie, its qubits and logic gates, can be characterized in detail \cite{Blume-Kohout2017-vy}. However, extrapolating whole-system performance from this data is infeasible or requires approximations, and even in the few-qubit setting these extrapolations are often inaccurate \cite{Proctor2020-ky}.

In this work we address this challenge by introducing an efficient and scalable protocol for verifying that an algorithmic circuit, $c$, can be executed with low error.
Our technique utilizes the concepts of motion reversal (\ie, a Loschmidt echo) \cite{Loschmidt_undated-uy, Emerson2007-am, Emerson2005-fd, Proctor2020-ky} and randomization \cite{Emerson2007-am} to efficiently check that $c$ has been implemented to a specified fidelity, using various \emph{mirror circuits} \cite{Proctor2020-ky} built from $c$. Our \emph{mirror circuit fidelity estimation} (MCFE) procedure does not rely on any special properties of the circuit being tested or quantum algorithm being implemented, except that the ideal evolution is unitary. Our procedure is efficient and scalable.
The complexity of the required classical computations scales linearly in the size of $c$, \ie, in $d \times n$ where $n$ is the number of qubits and $d$ is the number of layers of logic gates. Furthermore, the amount of data required to estimate the fidelity of $c$ to a specified relative precision is independent of $n$. MCFE therefore completely avoids the famous exponential scaling problems of state and process tomography. We argue that our protocol enables accurate estimation of fidelity under broad circumstances, and we demonstrate it in simulations through validation of the output of quantum approximate optimization algorithm (QAOA) \cite{Farhi2014-bt} circuits on up to 100 qubits.

\begin{figure*}[t!]
\includegraphics[width=18cm]{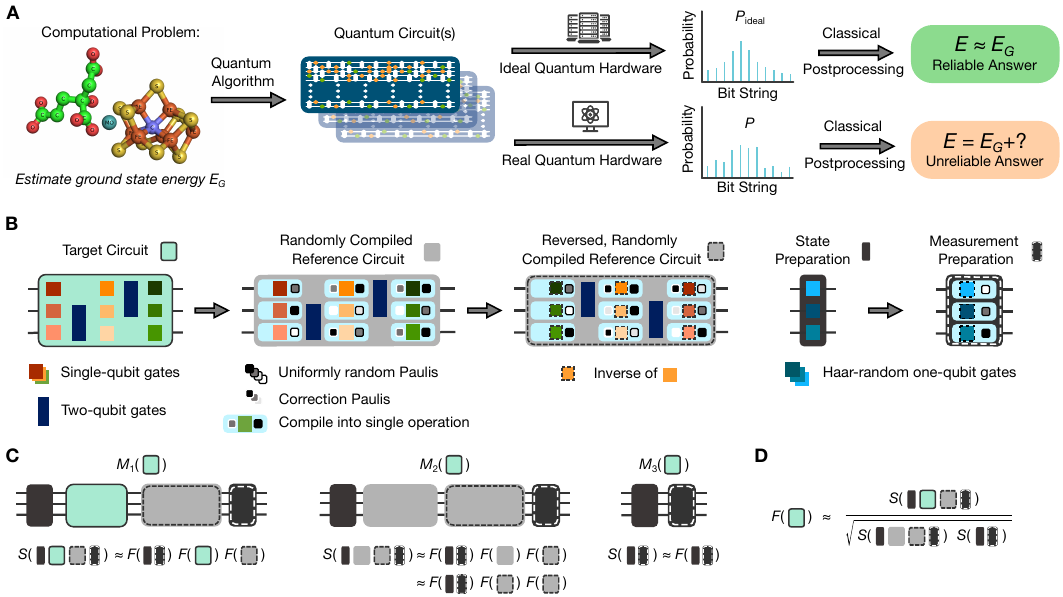} 
\caption{\textbf{Establishing trust in quantum computations by estimating circuit execution fidelity.} \textbf{A}. A quantum algorithm for gate-based quantum computers solves a computational problem by running one or more quantum circuits on a quantum computer, to sample from some probability distributions $P_{\textrm{ideal}}$. Real-world quantum computers implement circuits imperfectly, resulting in circuit outputs (bit strings) sampled from distributions $P$ that differ from $P_{\textrm{ideal}}$, causing inaccuracy in the algorithm's solution. Quantifying the solution inaccuracy caused by these imperfections is difficult, because, for example, computing $P_{\textrm{ideal}}$ is generally infeasible using even the most powerful supercomputers. \textbf{B}. 
Our technique estimates the fidelity $F(c)$ with which an as-built quantum computer can execute some target circuit $c$ (green box), using motion reversal circuits built from $c$ and the four reference (sub)circuits shown here. \textbf{C.} The three motion reversal circuits we use to estimate fidelity, $M_i(c)$, selectively use randomized compilation to isolate $F(c)$ without requiring that $c$ is randomly compiled. These motion reversal circuits are designed so that their mean adjusted success probabilities $S(\cdot)$ [see \eqref{eq:S}] are approximately equal to the product of the fidelities of their constituent subcircuits. This means that the simple function of their mean adjusted success probabilities shown in \textbf{D} can be used to estimate $F(c)$.}
\label{fig:circuits}
\end{figure*}

Existing techniques for verifying the output of quantum computers \cite{Ferracin2021-vh, Ferracin2019-ou, Bennink2020-wl, Flammia2011-qj, Pallister2018-qe, Mahadev_2018, Gheorghiu2019-pa, Gheorghiu2019-pa, Zhu2021-cs, Barz2013-uk} have complementary properties to ours. These methods can be divided into those that, like ours, quantify the accuracy with which a quantum circuit can be executed on a quantum computer that is noisy and error-prone but otherwise trusted, and methods for verifying the correct operation of a potentially malicious quantum computing server. Unlike our technique, the existing methods for quantifying the accuracy of circuit executions are either designed for special kinds of circuits---such as randomly compiled circuits \cite{Ferracin2021-vh, Ferracin2019-ou}, circuits that create anti-concentrated states \cite{Bennink2020-wl}, or Clifford circuits \cite{Pallister2018-qe}---or are only efficient for special kinds of circuits \cite{Flammia2011-qj, Leone2022-ls}. In particular, direct fidelity estimation \cite{Flammia2011-qj} is a technique for measuring the execution fidelity of any circuit, but it is only efficient for circuits that can be efficiently simulated classically \cite{Leone2022-ls}. The existing techniques for verifying the output of a potentially malicious quantum computer---using interactive proofs, blind quantum computing, and cryptographic methods---require resource-intensive implementations of algorithms that are experimentally challenging \cite{Barz2013-uk, Mahadev_2018, Gheorghiu2019-pa, Gheorghiu2019-pa, Zhu2021-cs, Stricker_2022}. It is unlikely that they can be used to establish trust in the first practically important quantum computations.

\section*{Results}
\subsection*{Quantum circuits}
An $n$-qubit quantum circuit $c$ is a sequence of $d$ layers of logic gates $l_1$, $l_2$, $\dots$, $l_d$, denoted by $c=l_d l_{d-1}\cdots l_2 l_1$, where a layer $l$ consists of gates applied to disjoint sets of the $n$ qubits (see examples in Fig.~\ref{fig:circuits}). In this work, a circuit or layer $l$ ideally implements an $n$-qubit unitary $U(l) \in \mathbb{SU}(2^n)$. Running a circuit $c$ consists of applying a state preparation that ideally initializes each of $n$ qubits in $\ket{0}$, then applying the layers of $c$ in turn, and then applying a measurement that ideally projects onto the computational basis. Our method assumes the ability to implement arbitrary single-qubit gates and an entangling two-qubit gate that is Clifford and self-inverse, \eg, CNOT. Note, however, that our technique can be applied to a circuit $c$ containing any gates, as we only require gates of this form in reference circuits against which $c$ is compared. We use $l_{\textrm{rev}}$ to denote the unique layer in which each gate in $l$ is replaced with its inverse, and $c_{\textrm{rev}} = l_{\textrm{rev},1}l_{\textrm{rev},2}\cdots l_{\textrm{rev},d}$ the motion reversal circuit, so $U(c_{\textrm{rev}})=U^{\dagger}(c)$ and $U(c_{\textrm{rev}}c)=\mathbb{I}$. The circuits used in our method will contain random layers and sub-circuits. A circuit containing random layers or subcircuits is a circuit-valued random variable, and is denoted using upper-case font and referred to as a randomized circuit. The expectation value, which is taken over all random variables within its argument, is denoted by $\mathbb{E}\{\cdot\}$.

\subsection*{Quantifying quantum computation accuracy}
There are many ways to quantify how well a quantum computer can run some $n$-qubit circuit $c$. Each run of $c$ generates an $n$-bit string sampled from some probability distribution $P$ over $\{0,1\}^n$. In the presence of errors, this is typically different from the distribution from which $c$ is supposed to sample ($P_{\textrm{ideal}}$), given by  
\begin{equation}
P_{\textrm{ideal},x}(c) = \bra{x} \; U(c)\ket{0}^{\otimes n},
\end{equation}
for $x \in \{0,1\}^n$.
One way to quantify how accurately a quantum computer can implement $c$ is by some measure of the distance between $P$ and $P_{\textrm{ideal}}$ (see Fig.~\ref{fig:circuits}A). However, this typically requires knowing $P_{\textrm{ideal}}$, and using a classical computer to calculate or sample from $P_{\textrm{ideal}}$ is infeasible for most circuits with $n \gg 1$, including all computationally useful circuits.

An alternative way to quantify how accurately $c$ is implemented is by some notion of evolution accuracy, \ie, the difference between the implemented evolution $\phi(c)$ and the ideal unitary $U(c)$. This is a strictly more rigorous test of a quantum computer's implementation of $c$, and it does not require a choice of input state (of which there are infinitely many possibilities). Testing evolution accuracy rather than comparing $P$ to $P_{\textrm{ideal}}$ is therefore important if $c$ is a subroutine, meaning that it is to be embedded within a variety of larger circuits. This is the case if $c$ prepares a state on which multiple non-commuting observables are to be measured, or if it is a multi-purpose computational primitive used within larger algorithms. To define notions of the evolution accuracy we require a mathematical model for $\phi(c)$. We use the standard ``Markovian'' model \cite{Nielsen2020-lt} in which $\phi(c)$ is a superoperator that maps $n$-qubit quantum states to $n$-qubit quantum states. We compare this to the superoperator representation of the ideal evolution $U(c)$, denoted by $\mathcal{U}(c)$ and defined by $\mathcal{U}(c)[\rho] = U(c)\rho U^{\dagger}(c)$. 

In this work we focus on estimating the \emph{entanglement fidelity} of $\phi(c)$ to $\mathcal{U}(c)$, defined by
\begin{equation}
    F(c) \equiv F(\mathcal{E}(c)) =  \langle \varphi | (\mathbb{I} \otimes \mathcal{E}(c))[|\varphi \rangle \langle \varphi |]|\varphi \rangle,
\end{equation}
where $\mathcal{E}(c)=\mathcal{U}^{\dagger}(c)\phi(c)$ is, be definition, $c$'s error map and $\varphi$ is any maximally entangled state of $2n$ qubits \cite{Schumacher1996-nv, Horodecki1999-rk, Nielsen2002-iu}. This fidelity is linearly related to another widely-used fidelity variant---the \emph{average gate fidelity} \cite{Horodecki1999-rk, Nielsen2002-iu}
\begin{align}
    \bar{F}(c) \equiv \bar{F}(\mathcal{E}(c)) &=  \int d \psi \, \bra{\psi}\mathcal{E}(c)[\psi] \ket{\psi}, \\
    &= \bra{\psi_0} \mathbb{E}\{ \mathcal{T}^{\dagger}  \mathcal{E}(c) \mathcal{T}\}[\psi_0] \ket{\psi_0} \label{eq:2design-twirl} ,\\
    & = \frac{2^n F(c) + 1}{2^n + 1}
\end{align}
where $d\psi$ is the Haar measure on pure states, $\psi_0$ is any pure state, and $\mathcal{T}$ is a superoperator-valued random variable with $\mathcal{T}[\rho]= U\rho U^{\dagger}$ where $U$ is a unitary 2-design over $\mathbb{SU}(2^n)$ \cite{Magesan2012-ca}.

\subsection*{Fidelity estimation using motion reversal} In this work, we show how a form of motion reversal circuits can be used for efficient and robust fidelity estimation. We first explain why motion reversal is an appealing method for estimating fidelity and why, in its simplest form, it fails to do so reliably. The reverse of $c$ ($c_{\textrm{rev}}$) ideally implements $\mathcal{U}^{\dagger}(c)$. So, because $\phi(c)=\mathcal{U}(c)\mathcal{E}(c)$ (as $\mathcal{E}(c)$ is \emph{defined} by $\mathcal{E}(c)=\mathcal{U}^{\dagger}(c)\phi(c)$), we can isolate $c$'s error map if we follow $c$ by an ideal reverse evolution. By interrogating this error map with sufficiently diverse states and measurements, we can then estimate $F(c)$. Equation (\ref{eq:2design-twirl}) tells us that the states generated by a unitary 2-design are sufficiently diverse. In particular, consider the circuit $M(c) = T_{\textrm{rev}}c_{\textrm{rev}}cT$, where $T$ is a randomized circuit such that $U(T)$ is a unitary 2-design over $\mathbb{SU}(2^n)$. Assume that all operations in $M(c)$ except $c$ are perfect, \ie, $\phi(c_{\textrm{rev}})=\mathcal{U}(c_{\textrm{rev}})$ etc. Then, by applying \eqref{eq:2design-twirl}, we find that these circuits' mean success probability $\bar{Q} = \mathbb{E}\{ P_0(M(c))\}$ satisfies $\bar{Q}=\bar{F}(c)$ and is therefore related to $F(c)$ by
\begin{equation}
   F(c) = (1 + \nicefrac{1}{2}^n)\bar{Q} - \nicefrac{1}{2^n}. \label{eq:F=Q}
\end{equation}

The problem with this fidelity estimation procedure is that no real quantum computer has access to perfect unitaries, state preparations, and measurements. Equation (\ref{eq:F=Q}) is not a reliable estimate of $F(c)$ because $\bar{Q}$ captures the impact of all errors in $M(c) = T_{\textrm{rev}}c_{\textrm{rev}}cT$, not just those in $c$. The error in $M(c)$ can be dominated by the error in the randomized state preparation and measurement subcircuits, $T$ and $T_{\textrm{rev}}$. This is because implementing a unitary 2-design requires a large circuit \cite{Dankert2009-yt, Aaronson2004-ab} (e.g., $\mathcal{O}(n^2/\log n)$ one- and two-qubit gates are necessary for a random $n$-qubit Clifford gate \cite{Aaronson2004-ab, Patel2008-dv}), and this makes fidelity estimation using a unitary 2-design infeasible on more than a few qubits \cite{Proctor2019-wp,Proctor2021-bq}. Furthermore, counter-intuitive effects can occur, \eg, coherent errors in $c_{\textrm{rev}}$ can even exactly cancel coherent errors in $c$ \cite{Proctor2020-ky} yielding $\bar{Q}=1$ when $F(c)\ll 1$. This is a well-known issue with simple motion reversal. 
We solve these problems using a streamlined randomized state preparation and measurement procedure, and randomized motion reversal, which we now introduce in turn.

\subsection*{Streamlined fidelity estimation using local randomized states}
We overcome the prohibitive size of the circuits required to implement an $n$-qubit unitary 2-design, by employing local randomized state preparation and measurement of each of the $n$ qubits, using ideas similar to those in early works on fidelity estimation and other methods based on randomized measurements \cite{Emerson2007-am, Brydges2019-nb, Flammia2011-qj}. The key insight is that an $n$-qubit unitary 2-design average can be mimicked using only single-qubit unitary 2-design averages, which require only single-qubit gates, and classical data processing. In particular, for any bit string $y$ and error map $\mathcal{E}$,
\begin{equation}
  F(\mathcal{E}) = \sum_{x} \left(-\nicefrac{1}{2}\right)^{h(x,y)} \bra{x}\mathbb{E}\{\mathcal{L}^{\dagger} \mathcal{E}  \mathcal{L} \}[\ket{y}\bra{y}]\ket{x},
  \label{eq:local-twirl-F}
\end{equation}
where $h(x,y)$ is the Hamming distance from $x$ to $y$, and $\mathcal{L} = \otimes_{i=1}^{n}\mathcal{L}_i$ where $\mathcal{L}_i[\rho]= U_i \rho U_i^{\dagger}$ and $U_i$ is a single-qubit unitary 2-design [see Supporting Information (SI)~\ref{app:fidelity} for a proof]. The summation over all bit strings, with the probability of each bit string multiplied by a Hamming distance dependent factor, accounts for the local nature of the randomized state preparation. This recovers the full power of an $n$-qubit unitary 2-design average using simple and efficient classical data processing. To use \eqref{eq:local-twirl-F} to estimate fidelity from running motion reversal circuits we compute their \emph{adjusted success probability}, defined by
\begin{equation}
    S(c) = \sum_{k=0}^n\left(-\frac{1}{2}\right)^{k}h_k(c).
    \label{eq:S}
\end{equation}
Here $h_k(c)$ is the probability that $c$ outputs a bit string $y$ whose Hamming distance from $c$'s target bit string $x_c$ is $k$, \ie, $h_k(c) = \sum_{y \in D_k} P_y(c)$ where $D_k =\{y \mid h(x_c,y)=k\}$. A circuit's target bit string $x_c$ is the bit string that satisfies $P_{\textrm{ideal},x_c}(c)=1$. Not all circuits have a target bit string, but all our motion reversal circuits do.

Equation~(\ref{eq:local-twirl-F}) implies that we can estimate fidelity using a motion reversal circuit that does not include full $n$-qubit unitary 2-design averaging, and its prohibitive overhead. Instead of using the motion reversal circuit $M(c)=T_{\textrm{rev}}c_{\textrm{rev}}cT$, we can estimate $F(c)$ by executing instances of
\begin{equation}
M'(c) = L_{\textrm{rev}}c_{\textrm{rev}}cL, \label{eq:m1}
\end{equation}
where $L$ is a randomized circuit such that $\mathcal{U}(L)=\mathcal{L}$ for $\mathcal{L}$ as stated above. We then estimate $\bar{S} = \mathbb{E}\{ S(M'(c))\}$, with $M'(c)$'s target bit string being simply the all-zeros string---as this circuit applies an identity operation if implemented without error. If all operations in $M'(c)$ except $c$ are perfect, \eqref{eq:local-twirl-F} implies that $F[\mathcal{E}(c)]= \bar{S}$. Although no operations can be implemented perfectly in practice, in streamlining our motion reversal circuit from $M(c)$ to $M'(c)$ we have constructed a circuit where the error in the randomized state preparation and measurement circuits will not dominate the error in $c$ in the $n\gg 1$ regime, for interesting $c$. To construct a reliable fidelity estimation method, we now only require a method for distinguishing errors in $c$ from those in other parts of our motion reversal circuits.

\begin{figure*}[t!]
\includegraphics[width=18cm]
{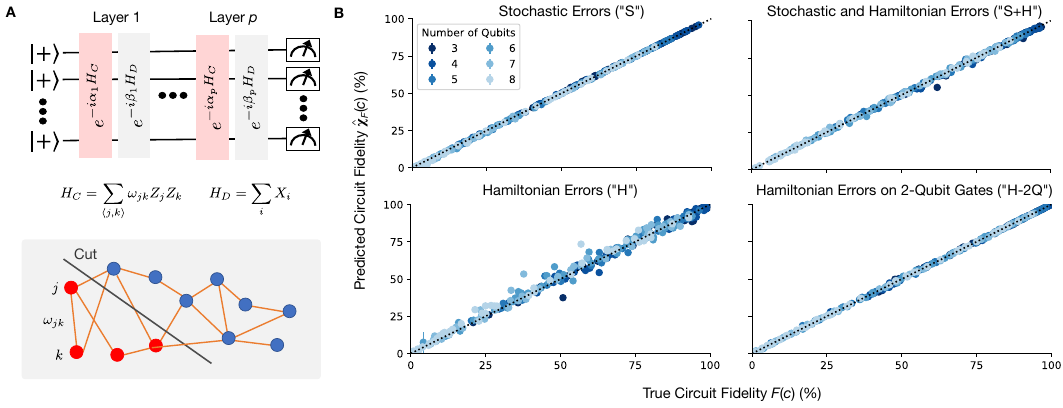}
\caption{\textbf{Applying our protocol to QAOA circuits using simulated data.} \textbf{A.} Parameterized QAOA circuits for approximately solving the MaxCut problem on a weighted $n$-vertex graph consist of alternating applications of unitaries generated by two Hamiltonians: $H_C$, which encodes the graph, and $H_D$, which is a driver Hamiltonian that does not commute with $H_C$ \cite{Farhi2014-bt}. These unitaries are compiled into single-qubit gates and CNOTs. We randomly selected 300 distinct QAOA circuits, on up to $n=8$ qubits and with up to $p=10$ algorithmic layers. We applied our protocol to each circuit, with simulated data from four different families of error models. The variational parameters were randomly assigned for each circuit. \textbf{B.} Each circuit's true process fidelity [$F(c)$], under the selected error model, versus the process fidelity predicted by our protocol [$\hat{\chi}_F(c)$], calculated by sampling 1000 mirror circuits of each type (see Fig.~\ref{fig:circuits}C) and simulating them under that error model. The four error models consist of stochastic Pauli errors on all gates, stochastic Pauli errors and Hamiltonian errors on all gates, Hamiltonian errors on all gates, and Hamiltonian errors on only two-qubit gates. The rates of the errors in each error model are randomly sampled for each QAOA circuit $c$.  Each $\hat{\chi}_F(c)$ is shown with error bars (1 standard deviation, calculated using a nonparametetric boostrap) that, in most cases, are smaller than the data points.}
\label{fig:simulations}
\end{figure*}

\subsection*{Randomized motion reversal}
Our aim is to robustly estimate $F(c)$, the fidelity of $c$'s overall error map $\mathcal{E}(c)$. Errors on the individual gates and layers of $c$ may coherently add or cancel, and this is fine---we wish to quantify how they impact the fidelity with which $c$ is implemented. However, fidelity estimation requires embedding $c$ within larger circuits. Now $c$'s errors could coherently add or cancel with errors in other parts of these larger circuits, \eg, $c_{\textrm{rev}}$. We need to prevent coherent addition or cancellation of errors between $c$ and other subcircuits without changing error propagation processes within $c$. To achieve this, we use selective \emph{randomized compilation} \cite{Wallman2016-rd, Knill2005-xm}. A randomly compiled circuit has only stochastic errors, which cannot coherently add. In particular, we will compare $c$ to a randomly compiled motion reversal circuit. Unlike techniques designed to predict the fidelity of randomly compiled circuits \cite{Ferracin2021-vh, Ferracin2019-ou, Flammia2021-dn, Harper2020-te, Flammia2020-ym, Erhard2019-wk}, our technique runs $c$ without randomized compilation. It uses randomly compiled circuits only to provide a stable reference frame against which $c$ can be compared.

Randomized compilation is defined on circuits with a particular structure, but we do not require $c$ to have that structure. So to implement randomized compilation on a circuit $c$, we first construct a circuit
\begin{equation}
    \tilde{c} = l_{\tilde{d}}e_{\tilde{d}-1}l_{\tilde{d}-1} \cdots e_2 l_2 e_1 l_1,
    \label{eq:alternating-circuit}
\end{equation}
that is logically equivalent to $c$, meaning that  $\mathcal{U}(\tilde{c}) = \mathcal{U}(c)$, where each $e_i$ layer contains only two-qubit gates that are Clifford and self-inverse, and each $l_i$ layer contains only single-qubit gates. Layers are allowed to be empty. Constructing a circuit of this form is always possible, as the set of all single-qubit gates along with any entangling gate forms a universal gate set \cite{Brylinski2002-tl}, and it is efficient for any circuit in which all gates interact $k$ or fewer qubits for some constant $k$ (e.g., any circuit containing only one- and two-qubit gates). Note that $c$ may already have this form, in which case we can set $\tilde{c}=c$. This is the case in our simulations (reported below) and in the schematic of our method in Fig.~\ref{fig:circuits}B. 

For a circuit $\tilde{c}$ with the form of \eqref{eq:alternating-circuit}, a randomized compilation of $\tilde{c}$, labeled $f_{\textrm{rc}}(\tilde{c})$, is constructed by first sampling $\tilde d$ $n$-qubit Pauli superoperators, $\mathcal{P}_i$ for $i=1,2,\dots, \tilde d$. We then replace each layer $l_i$ in $\tilde c$ with a new layer $l^\prime_i$ satisfying $\mathcal{U}(l^\prime_i) = \mathcal{P}_i \mathcal{U}(l_i) \mathcal{U}(e_{i-1})\mathcal{P}_{i-1}\mathcal{U}(e_{i-1})^\dagger$, with $\mathcal{P}_0$ defined to be the identity operator and $e_0$ an empty layer. This new circuit implements the same unitary as $\tilde c$ up to post-multiplication by a Pauli operator $\mathcal{P}_{\tilde d}$. In our context, $\mathcal{P}_{\tilde d}$ simply changes the circuit's target bit string.

\subsection*{Robust mirror circuit fidelity estimation}
Our MCFE protocol combines motion reversal with randomized compilation and streamlined, local randomized state preparation to robustly estimate $F(c)$. MCFE uses the following three randomized circuits
\begin{align}
    M_{1}(c) &= f_{\textrm{rc}}(L_{\textrm{rev}}\tilde{c}_{\textrm{rev}})cL, \label{eq:M1}\\
    M_{2}(c) &= f_{\textrm{rc}}(L_{\textrm{rev}}\tilde{c}_{\textrm{rev}}\tilde{c}L), \label{eq:M2} \\
    M_{3}(c) &= f_{\textrm{rc}}(L_{\textrm{rev}}L),
    \label{eq:M3}
\end{align}
shown in Fig.~\ref{fig:circuits}B. To estimate $F(c)$ from running these circuits, MCFE uses the following formula:
\begin{equation} 
\chi_F(c) = 1 - \frac{4^n -1}{4^n}\left(1 - \frac{\mathbb{E}\{\gamma(M_1(c))\} }{\sqrt{\mathbb{E} \{\gamma(M_2(c))\} \mathbb{E} \{\gamma(M_3(c))\} }}\right), \label{eq:chiF}
\end{equation}
where $\gamma(c)$ is a circuit's \emph{effective polarization} \cite{Proctor2021-bq}
\begin{equation}
\gamma(c) =  \frac{4^n}{4^n-1}S(c) - \frac{1}{4^n -1}.
\label{eq:gamma}
\end{equation}

MCFE is the following procedure, given a target circuit $c$: 
\begin{enumerate}
    \item Sample $N_i \gg 1$ of the $M_i(c)$ mirror circuits, with $i=1,2,3$, and execute each one $K \geq 1$ times.
    \item Estimate $c$'s process fidelity by computing $\chi_F(c)$ from the data, where each average in $\chi_F(c)$ is estimated using the sample mean.
\end{enumerate}
In SI~\ref{app:theory} we show that 
\begin{equation}
    \chi_F(c) = F(c) + \mathcal{O}(\Delta_1, \Delta_2), \label{eq:chi-F-maintext}
\end{equation}
where $\Delta_1$ and $\Delta_2$ are (typically) small error terms whose form is provided in SI~\ref{app:theory}. That theory assumes: Markovian errors (as assumed throughout the main text); errors on single-qubit gates are unchanged under Pauli randomization (gate-independence of these errors is sufficient); and a simplified model for state-preparation and measurement errors. See SI~\ref{app:theory} for precise statements of these assumptions.

We now explain the reasoning behind the MCFE protocol, i.e., why we should expect \eqref{eq:chi-F-maintext} to hold. The three familes of mirror circuit used in MCFE combine the motion-reversal circuit of \eqref{eq:m1} with randomized compilation. Errors within the randomly compiled components of $M_i(c)$ become purely stochastic after averaging \cite{Wallman2016-rd, Knill2005-xm, Hashim2023-qk}. This enables separating out the errors due to the different subcircuits within $M_i(c)$. This is because the overall fidelity of a sequence of stochastic error channels is the product of their fidelities, to first order in their infidelities. Similarly, the fidelity of any error channel---such as $\mathcal{E}(c)$---followed by a stochastic error channel is equal to the product of the fidelities of these two channels, to first order in their infidelities. This approximate factoring of fidelities is, in our view, the key reason why MCFE works (i.e., why $\Delta_1$ and $\Delta_2$ in \eqref{eq:chi-F-maintext} are small), and we justify this approximation in detail in SI~\ref{app:theory}, as part of our theory of MCFE.
In the context of MCFE, this implies that $S_i \equiv S(\mathbb{E}\{M_i(c)\})$ is approximately the product of the fidelities of $M_i(c)$'s constituent subcircuits. In particular,
\begin{align}
   S_1 &\approx F_{0}F(\tilde{c}_{\textrm{rev}})F(c) \label{eq:s1toF},\\
   S_2 &\approx F_{0}F(\tilde{c}_{\textrm{rev}}) F(\tilde{c})\label{eq:s2toF} ,\\
   S_3 &\approx F_{0}, \label{eq:s3toF}
\end{align}
where $F_0$ is the fidelity of randomized state preparation and measurement operations (including contributions from errors in $L$ and $L_{\textrm{rev}}$, errors in the readout, and errors in preparing the standard input state). Furthermore, we have that 
\begin{equation}
F(\tilde{c}_{\textrm{rev}}) \approx F(\tilde{c}) \label{eq:FcrevtoFc}
\end{equation}
 as both circuits are randomly compiled and contain identical two-qubit gate layers. Therefore, as shown diagrammatically in Fig.~\ref{fig:circuits}C, we can combine the four approximate equalities of Eqs.~(\ref{eq:s1toF})-(\ref{eq:FcrevtoFc}) to compute $F(c)$:
\begin{equation}
F(c) \approx \frac{S_1}{\sqrt{S_2S_3}}. \label{eq:F(c)-simple}
\end{equation}

For $n \sim 1$ we get a better approximation by using \emph{polarization} rather than fidelity. 
That is why we estimate $F(c)$ using \eqref{eq:chiF} rather than the simpler expression of \eqref{eq:F(c)-simple}, which differ by a negligable factor when $n \gg 1$, as $\chi_F(c) = \nicefrac{S_1}{\sqrt{S_2S_3}} + \mathcal{O}(\nicefrac{1}{4^n})$. 

\subsection*{Sample complexity}
Our formula for $\chi_F(c)$ includes expectation values over large circuit ensembles, and we must estimate $\chi_F(c)$ from a finite amount of data (finite $N_i$ and $K$ in the MCFE procedure). This estimation can be done efficiently, because the required total number of samples $K_{\textrm{total}} = (N_1+N_2+N_3)K$ to estimate $\chi_F(c)$ to a specified relative precision does not grow with $n$. In SI~\ref{sec:resources}, we show that 
\begin{equation}
    \label{eq:N}
    N_i = \mathcal{O}\left(\frac{\ln(1/\delta)}{\alpha^2\gamma_{\min}^2} \right)
\end{equation}
circuits are sufficient to estimate $\chi_F(c)$ to relative precision $2\alpha$, with probability greater than $(1-\delta)^3$, 
where $\gamma_{\min}$ is a constant such that $\mathbb{E}\{\gamma(M_i(c))\} > \gamma_{\min}$ for $i=1,2,3$. Our protocol can therefore be efficiently applied in the $n \gg 1$ regime where direct verification of a quantum computation's accuracy, via classical simulations, is intractable, as long as $\gamma_{\min}$ is not too small. Note, however, that, for a given quantum computer (or a realistic error model of such a system), $\gamma_{\min}$ will typically decrease with increasing $n$---and so, with a given system, the magnitude of its error will place a limit on the circuits for which MCFE is feasible. 

\subsection*{Simulations}
We tested our method in simulations, by applying it to circuits that implement QAOA for the MaxCut problem on $n$-vertex weighted graphs \cite{Farhi2014-bt}. QAOA approximately solves MaxCut using the parameterized circuit shown in Fig.~\ref{fig:simulations}A, which is parameterized by the problem graph, the number of algorithmic layers $p$, and two length $p$ vectors of angles $\vec{\alpha}, \vec{\beta} \in (-\pi,\pi]^{p}$. We applied our method to 300 distinct QAOA circuits, obtained by varying $n$ and $p$ ($n=3,4,\dots,8$ and $p=1,2,5,8,10$), and at each pair $(n,p)$ sampling 10 distinct $n$-vertex weighted graphs and random angle vectors $\vec{\alpha}, \vec{\beta}$. For each QAOA circuit $c$, we sample $N$ of each of the three types of mirror circuit, and we simulate all $3N+1$ circuits under each of four different error models (detailed below). The main purpose of these simulations is to compare $\chi_F(c)$ to $F(c)$---as we do not expect $\chi_F(c)$ to exactly equal $F(c)$---rather than to study the impact of finite $N$ and $K$ on our estimate's statistical uncertainty. We therefore sample many mirror circuits ($N=1000$) and compute the exact $\gamma$ for each circuit (equivalent to $K\to\infty$), as then finite-sample fluctuations will be small, \ie, $|\hat{\chi}_F(c) -\chi_F(c)| \ll 1$, and they will not obscure any discrepancies between $\chi_F(c)$ and $F(c)$.

Figure~\ref{fig:simulations}B shows $F(c)$ versus $\hat{\chi}_F(c)$ for all 300 QAOA circuits, with each of the scatter plots showing results for a different family of error models. These error models consist of stochastic Pauli errors on all gates (``S''), stochastic Pauli errors and Hamiltonian (\ie, coherent) errors on all gates (``S+H''), Hamiltonian errors on all gates (``H''), and Hamiltonian errors on only two-qubit gates (``H-2Q'') \cite{blumekohout_2021-tr}. The rates of the errors in each error model are randomly sampled for each QAOA circuit $c$, so each data point in Fig.~\ref{fig:simulations}B compares $\chi_F(c)$ and $F(c)$ for a unique QAOA circuit $c$ simulated under a unique error model. See SI~\ref{app:simulations} for details of these error models and simulations.

We find that $\chi_F(c)$ is a very good approximation to $F(c)$ under S, S+H, and H-2Q error models, whereas $\chi_F(c)$ differs from $F(c)$ more substantially for the H models. For S models the estimation error is particularly small, \eg, $|[\hat{\chi}_F(c) - F(c)]/F(c)| \leq 0.4\%$ for all circuits and S models in which $F(c) \geq 75\%$. The larger estimation error when there are Hamiltonian errors on both one- and two-qubit gates is consistent with our theory for our protocol---as applying our theory to these error models requires larger approximations. In particular, general Hamiltonian errors on one-qubit gates introduces approximation into the theory of randomized compiling that we rely on to show that $\chi_F(c)\approx F(c)$ (specifically, this model violates our assumption that a single-qubit gate's error channel is unchanged when that gate is Pauli randomized). See SI~\ref{app:theory} for further discussions.

\begin{figure}[t!]
\includegraphics[width=8.5cm]{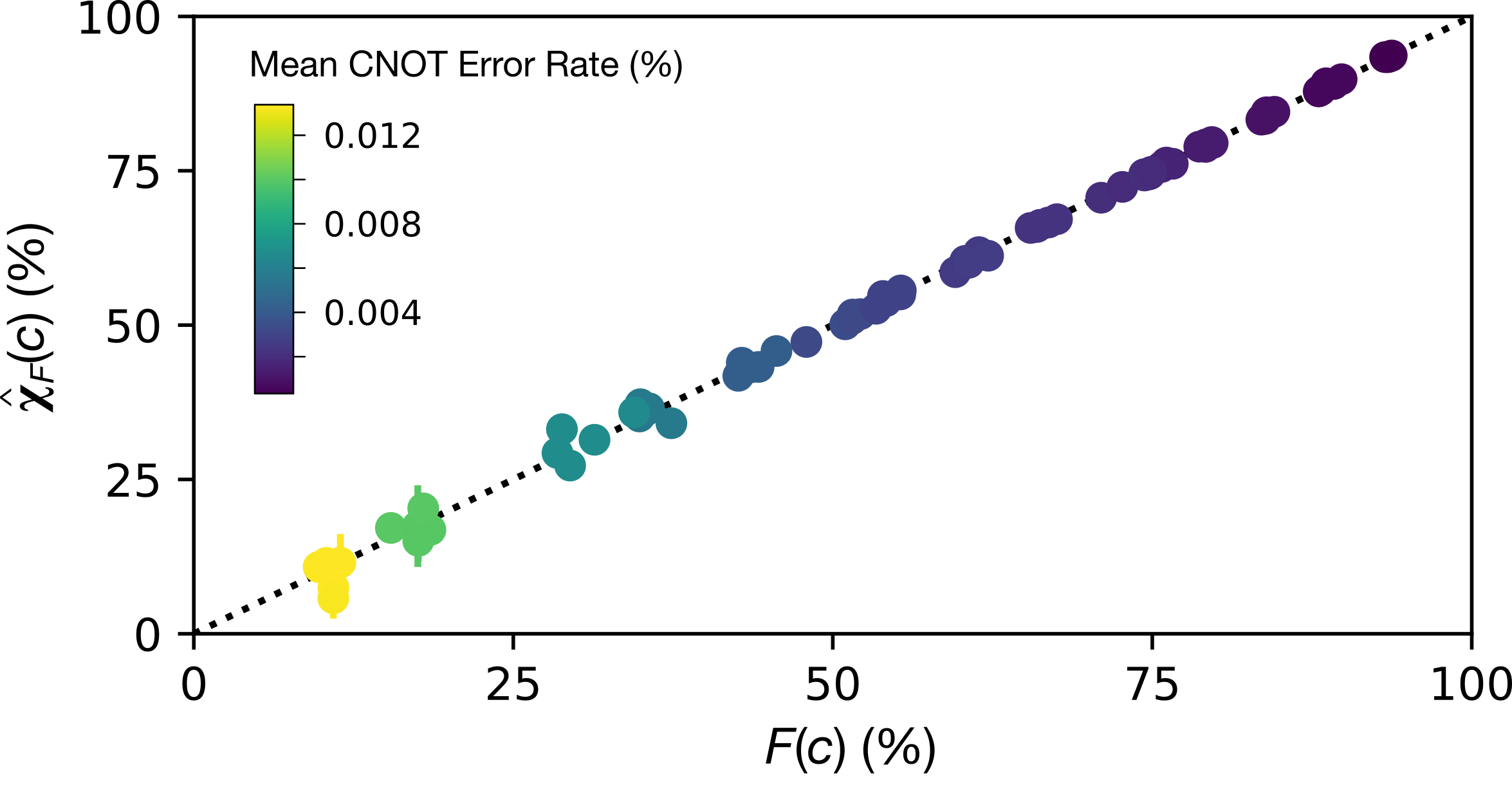}
\caption{\textbf{Measuring the fidelity of 100-qubit circuits.} We simulated applying our technique to 100-qubit circuits, demonstrating that it works in the $n=\mathcal{O}(100)$ regime of potential quantum advantage, using only 1,200 circuits and a total of 216,000 samples for each $c$. These simulations use Clifford circuits subject to stochastic Pauli errors so that we can efficiently generate simulated data. For every selected circuit and error model (details in text), we find that the estimate of the fidelity obtained from our protocol [$\hat{\chi}_F(c)$] is a good approximation to the true fidelity [$F(c)$]. Each $\hat{\chi}_F(c)$ is shown with error bars (1 standard deviation, calculated using a nonparametetric boostrap) that, in most cases, are smaller than the data points.}
\label{fig:simulations-2}
\end{figure}

Our method can be efficiently applied to general $n$-qubit circuits for any $n$, but we cannot generate simulated data for general circuits with $n\gg 1$. So, to investigate its performance in the $n\gg 1$ regime, we constructed 75 100-qubit, $p=1$ QAOA circuits for unweighted graphs with angles restricted to $\vec{\alpha},\vec{\beta} \in \{-\nicefrac{\pi}{2}, 0,\nicefrac{\pi}{2},\pi\}^p$ and applied our protocol to data obtained from simulating these circuits in the presence of stochastic Pauli errors (see SI~\ref{app:simulations} for details). This restriction on the QAOA angles means that these circuits are all Clifford circuits, so these simulations are efficient in $n$ \cite{Aaronson2004-ab}. We find that the estimate $\hat{\chi}_F(c)$ of the fidelity obtained from our protocol is a very good approximation to the true fidelity $F(c)$, while using only 1,200 circuits and 216,000 total samples for each circuit $c$ ($K=180$, $N=400$). This demonstrates that our protocol can be applied in the $\mathcal{O}(100)$ qubit regime of potential quantum advantage \cite{Preskill2018-jz, Farhi2014-bt, Bharti2022-rz, Guerreschi2019-mg, Nam_2020, Yeter-Aydeniz_2020, Tazhigulov_2022}.

\section*{Discussion}
State-of-the-art quantum computers can no longer be simulated on classical supercomputers. This makes it impossible to directly quantify the accuracy of their output when running algorithms that do not have efficiently verifiable solutions. In this paper we have introduced an efficient and simple technique for estimating the fidelity with which a quantum circuit $c$ can be executed on a specific quantum computer. Our technique can be used to confirm that a full-scale quantum computation has been implemented with low error, even using hardware that suffers complex and not fully understood errors. Assessing the accuracy of a complex experiment and ascribing confidence in its output is a familiar challenge in experimental science, and our technique is a general solution to this problem for quantum computing systems.

The immediate and near-term use of our protocol is for quantifying the accuracy of NISQ (noisy intermediate scale quantum) computations. However, it will still be necessary to quantify circuit execution accuracy in the era of fault-tolerant quantum computation---because quantum computations on error corrected qubits will still suffer errors at the logical level, albeit at suppressed rates.
We anticipate that the execution of our protocol at the logical level will enable efficient verification of future fault-tolerant quantum algorithmic circuits. Our method will therefore enable trustworthy quantum computation in both the NISQ computing era and beyond.

\begin{acknowledgements}
This work was supported by the U.S. Department of Energy, Office of Science, Office of Advanced Scientific Computing Research through the Accelerated Research in Quantum Computing (ARQC) program, the Quantum Computing Application Teams (QCAT) program, and the Quantum Testbed Program, and the Laboratory Directed Research and Development program at Sandia National Laboratories. Sandia National Laboratories is a multi-program laboratory managed and operated by National Technology and Engineering Solutions of Sandia, LLC., a wholly owned subsidiary of Honeywell International, Inc., for the U.S. Department of Energy's National Nuclear Security Administration under contract DE-NA-0003525. All statements of fact, opinion or conclusions contained herein are those of the authors and should not be construed as representing the official views or policies of the U.S. Department of Energy, or the U.S. Government.
\end{acknowledgements}

\bibliography{Bibliography}

\section*{Supporting Information}

\subsection{Computing fidelity using local twirls}\label{app:fidelity}

In this section we prove \eqref{eq:local-twirl-F}, where we claim that for any error map $\mathcal{E}$, $F(\mathcal{E})$ is equal to the quantity
\begin{equation}
  \Upsilon  \equiv \sum_{x} \left(-\nicefrac{1}{2}\right)^{h(x,y)} \bra{x}\mathbb{E}\{\mathcal{L}^{\dagger} \mathcal{E}  \mathcal{L}\}[\ket{y}\bra{y}]\ket{x},
\end{equation}
where $y$ is any $n$-bit string, $h(x,y)$ is the Hamming distance from $x$ to $y$, and 
\begin{equation}
   \mathcal{L} = \otimes_{i=1}^{n}\mathcal{L}_i.
\end{equation}
Here $\mathcal{L}_i[\rho]= L_i \rho L_i^{\dagger}$ and $L_i$ is a single-qubit unitary 2-design. It is useful to first rewrite $\Upsilon$ as
\begin{equation}
  \Upsilon =\sum_{k=0}^n \left(-\nicefrac{1}{2}\right)^{k} h_k(\mathcal{E}),
\end{equation}
where 
\begin{equation}
    h_k(\mathcal{E}) = \sum_{h(x,y)=k} \bra{x}\mathbb{E}\{\mathcal{L}^{\dagger} \mathcal{E}  \mathcal{L}\}[\ket{y}\bra{y}]\ket{x},
\end{equation}
is the probability of observing any bit string $x$ that is a Hamming distance of $k$ from $y$.

We begin by noting that twirling $\mathcal{E}$ by a tensor product of single-qubit 2-design projects $\mathcal{E}$ onto the $2^n$ dimensional space spanned by tensor products of single-qubit depolarizing channels \cite{Gambetta2012-zd}. That is
\begin{equation}
    \bar{\mathcal{E}} = \mathbb{E}\{\mathcal{L}^{\dagger} \mathcal{E}  \mathcal{L}\}
\end{equation}
is a stochastic Pauli channel with (1) equal \emph{marginal} probabilities to induce an $X$, $Y$ or $Z$ error on any fixed qubit, and (2) a fidelity equal to that of $\mathcal{E}$, \ie, 
\begin{equation}
   F(\bar{\mathcal{E}})=F(\mathcal{E}). 
   \label{eq:equal_fid}
\end{equation}
Therefore, in the following we prove that $\Upsilon = F(\bar{\mathcal{E}})$, from which \eqref{eq:local-twirl-F} follows immediately.

The channel $\bar{\mathcal{E}}$ is entirely described by a set of $2^n$ probabilities $\{r_{Q}(\bar{\mathcal{E}})\}$ where $Q$ runs over all possible subsets of the $n$ qubits and $r_Q(\bar{\mathcal{E}})$ is the probability that $\bar{\mathcal{E}}$ applies a nontrivial Pauli error to those qubits---meaning that $\bar{\mathcal{E}}$ applies a Pauli that is not the identity on each qubit in $Q$ and that is the identity on all other qubits. Conditioned on $\bar{\mathcal{E}}$ applying an error to the qubit set $Q$, the error that occurs is a uniformly random sample from the set of $3^{W(Q)}$ possible $n$-qubit Pauli operators that are the identity on all but the qubits in $Q$, where $W(Q)$ denotes the size of the set $Q$. Therefore, if an error occurs on the qubit set $Q$, which is a weight $w=W(Q)$ error, its action on a computational basis state $\ket{y}\bra{y}$ is to flip $k$ of the bits with probability 
\begin{equation}
A_{kw}= {w \choose k} \frac{2^k}{3^w},
\end{equation}
when $k\leq w$ and $A_{kw}=0$ otherwise. This is because the number of bits flipped corresponds to the number of $X$ and $Y$ errors in the applied Pauli error, and $A_{kw}$ is the probability that a uniformly random weight $w$ Pauli contains $(w-k)$ $Z$ operators. 

The quantity $h_k(\mathcal{E})$ is equal to the total probability of $\bar{\mathcal{E}}$ causing $k$ bit flips on $y$. This can be computed by summing, over all qubit subsets $Q$, the probability $r_{Q}(\bar{\mathcal{E}})$ of an error occurring on qubit set $Q$ multiplied by the probability that an error on that subset causes $k$ bit flips, which is $A_{kW(Q)}$. That is
\begin{align}
    h_{k}(\mathcal{E}) &= 
  \sum_{Q} A_{k W(Q)} r_{Q}(\bar{\mathcal{E}}),\\ 
  &= \sum_{w} A_{k w} r_{w}(\bar{\mathcal{E}})
\end{align}
where $r_w(\bar{\mathcal{E}}) = \sum_{Q \mid W(Q)=w} r_{Q}(\bar{\mathcal{E}})$ is the probability that $\bar{\mathcal{E}}$ applies any weight $w$ Pauli. This equation can be written as 
\begin{align}
    \vec{h}(\mathcal{E}) = A \vec{r}(\bar{\mathcal{E}}).
\end{align}

The fidelity of $\bar{\mathcal{E}}$ is the probability that it applies the identity Pauli, so 
$F(\bar{\mathcal{E}}) = r_0(\bar{\mathcal{E}})$, and therefore,
\begin{align}
 F(\bar{\mathcal{E}}) = [A^{-1}\vec{h}(\mathcal{E})]_0.
 \label{eq:FEbar_mat}
\end{align}
To obtain our result we only need the first row of the inverse of $A$, which is given by
\begin{equation}
    [A^{-1}]_{0k} = \left(-\nicefrac{1}{2}\right)^k.
\end{equation}
Substituting this into \eqref{eq:FEbar_mat} yields $F(\bar{\mathcal{E}})=\Upsilon$, implying $F(\mathcal{E})=\Upsilon$, by \eqref{eq:equal_fid}.

\subsection{Resource requirements}
\label{sec:resources}
Our procedure approximates the fidelity $F(c)$ of a quantum circuit, $c$, by estimating the quantity $\chi_F(c)$, which is defined in \eqref{eq:chiF}. In this section, we derive \eqref{eq:N}, which defines the number of samples (experimental runs of quantum circuits) needed to guarantee that our estimate is accurate to a specified relative error, with high probability. For the purposes of this section, we assume each circuit is run only $K=1$ times. 

For any given quantum circuit, $c$, we estimate $\chi_F(c)$ by first estimating the expected effective polarization 
\begin{equation}
    \gamma_i = \mathbb{E}\{\gamma(M_i(c))\}
\end{equation}
for each of three ensembles of mirror circuits, $M_i(c)$ with $i=1,2,3$. For each ensemble $M_i(c)$, we draw $N$ random circuits, run each circuit $K=1$ times on the quantum computer, determine the Hamming distance ($\{h_j^{(i)}\}_{j=1}^N$) of each resulting bit string to the circuit's target bit string, and compute the quantity:
\begin{equation}
    \hat\gamma_i = \sum_{j=1}^N \frac{1}{N} \left(
        \frac{4^n}{4^n-1}\left(-\frac{1}{2}\right)^{h_j^{(i)}} - \;\frac{1}{4^n-1}
        \right)
\end{equation}
The quantity $\hat\gamma_i$ is an unbiased estimate of the expected polarization, as its expected value  $\mathbb{E}\{\hat\gamma_i\}$ is equal to the true (unknown) expected polarization $\gamma_i$. Because each circuit is run only once, $\hat\gamma_i$ is a sum of bounded, independent random variables, and so obeys Hoeffding's concentration inequality \cite{Hoeffding1963-lq}:
\begin{equation}
\label{eq:gammaHoeffding}
    P\big(\vert \hat\gamma_i - \gamma_i \vert \ge \epsilon\big) \le 
        2 \exp\left(- \frac{8}{9}\left(\frac{4^n-1}{4^n}\right)^2 \epsilon^2 N \right).
\end{equation}

If we are assured that the mean effective polarizations of each of the circuit ensembles $\gamma_i$ is greater than some minimum, $\gamma_{\textrm{min}}$, then we can replace the additive error in \eqref{eq:gammaHoeffding} with a relative error. For any $\alpha >0$:
\begin{equation}
    P\big(\left\vert \hat\gamma_i - \gamma_i \right\vert \ge \alpha\,\gamma_i \big) \le \delta 
\end{equation}
where
\begin{equation}
    \label{eq:gammaHoeffdingRelativeError}
    \delta = 2\exp\left(- \frac{8}{9}\left(\frac{4^n-1}{4^n}\right)^2 \alpha^2\gamma_{\min}^2 N \right).
\end{equation}

The fidelity of circuit $c$ is estimated using \eqref{eq:chiF}, which we reproduce here written in terms of $\gamma_i$:
\begin{equation*}
\chi_F(c) = 1 - \frac{4^n -1}{4^n}\left(1 - \frac{\gamma_1 }{\sqrt{\gamma_2\gamma_3}}\right).
\end{equation*}
Using the bound above for the relative error in our estimates $\hat\gamma_i$, we can bound the error in our estimate of $\chi_F(c)$. The estimate of $\chi_F(c)$ depends linearly on the estimate of the ratio $R(c)$, defined as:
\begin{equation}
    \hat R(c) = \frac{\hat\gamma_1 }{\sqrt{\hat\gamma_2\hat\gamma_3}}. 
\end{equation}
The true value of this ratio is given in terms of the expectation values of $\gamma(M_i(c))$. 
\begin{equation}
    R(c) = \frac{\gamma_1}{\sqrt{\gamma_2\gamma_3}}.
\end{equation}
With probability greater than $(1-\delta)^3$, all three effective polarization estimates, $\hat\gamma_i$, will have relative error less than $\alpha$. Within this range, the worst case error on the estimate of $\hat R(c)$ occurs when $\hat\gamma_1$ has relative error $(1\pm\alpha)$ , and both $\hat\gamma_2$ and $\hat\gamma_3$ have relative errors $(1\mp\alpha)$.  
\begin{align}
    R^{\pm}(c) &= \frac{\gamma_1 (1\pm\alpha) }{\sqrt{\gamma_2(1\mp\alpha)\gamma_3(1\mp\alpha)}}\\
        &=R(c) \frac{1\pm\alpha}{1\mp\alpha}, \\
        &\simeq R(c) (1\pm2\alpha).
\end{align}
In the limit of small $\alpha$, we then have 
\begin{equation}
    P\Big( \vert \hat R(c) - R(c) \vert \le 2 \alpha\; R(c)  \Big) \ge (1-\delta)^3.
\end{equation}
Similarly, our estimate $\hat\chi_F(c)$ satisfies:
\begin{equation}
    P\Big( \vert \hat\chi_F(c) - \chi_F(c) \vert \le 2 \alpha \;\chi_F(c)  \Big) \ge (1-\delta)^3.
\end{equation}
For any fixed value of $\delta$, we can solve for $N$, the minimum number of circuits that must be sampled for each ensemble $M_i(c)$ to achieve this bound: 
\begin{equation}
    N \ge \frac{9}{8}\left(\frac{4^n}{4^n-1}\right)^2 \frac{\ln(2/\delta)}{\alpha^2\gamma_{\min}^2}.
\end{equation}
This is weakly decreasing in the number of qubits $n$. We can remove the dependence on $n$ by taking the limit of a single qubit. In this case, we have:
\begin{equation}
    N \ge \frac{2 \ln(2/\delta)}{\alpha^2\gamma_{\min}^2},
\end{equation}
which implies \eqref{eq:N} stated in the main text.

\subsection{Relating $\chi_F(c)$ to fidelity}\label{app:theory}
In this section we provide a theory that shows that $\chi_F(c) \approx F(c)$ under certain assumptions (stated below), and why we conjecture this holds under broad, physically relevant situations that go beyond our proof.
First, we define the \emph{polarization} of an error map $\mathcal{E}$, which is used extensively in our theory:
    \begin{equation}
        \lambda(\mathcal{E}) = \frac{4^n}{4^n - 1} F(\mathcal{E})  - \frac{1}{4^n - 1}.
    \label{eq:polarization}
    \end{equation}
Note that $\lambda(\mathcal{E}) = F(\mathcal{E}) + \mathcal{O}(\nicefrac{1}{4^n})$, so the difference between polarization and fidelity is negligible for $n\gg 1$. Our circuit ensembles, $M_i(c)$ with $i=1,2,3$, are designed so that 
\begin{equation}
\gamma_i \equiv \mathbb{E}\{\gamma(M_i(c))\}
\end{equation}
is approximately equal to the product of the polarizations of the constituent subcircuits in the $M_i(c)$ circuits. More precisely, we claim that
\begin{align}
   \gamma_1 & = \lambda_{0}\lambda_{\textrm{rev}} \lambda(c) + \Delta_1, \label{eq:gamma1}\\
   \gamma_2 & = \lambda_{0}\lambda_{\textrm{rev}}^2 + \Delta_2 \label{eq:gamma2} \\
   \gamma_3 &= \lambda_{0}, \label{eq:gamma3}
\end{align}
where $\Delta_1$ and $\Delta_2$ (defined later) are small, $\lambda_{0}$ is related to the mean polarization (or fidelity) of the randomized state preparation and measurement, $\lambda_{\textrm{rev}}$ is related to the mean polarization (or fidelity) of a random compilation of $\tilde{c}_{\textrm{rev}}$, and $\lambda(c)$ is the polarization of $\mathcal{E}(c)$.
If these three equations hold for small $\Delta_1$ and $\Delta_2$, then
\begin{equation}
\frac{\gamma_1}{\sqrt{\gamma_2\gamma_3}}  = \lambda(c) +\mathcal{O}(\Delta_1, \Delta_2),
\end{equation}
and therefore 
\begin{equation}
\chi_F(c)  =  F(c)  + \mathcal{O}(\Delta_1, \Delta_2).\label{eq:chi2-F-theory}
\end{equation}
Note that in the main text we state that $S_i\equiv \mathbb{E}\{S(M_i(c))\}$ is approximately equal to the product of the fidelities of the constituent subcircuits, rather than that the effective polarizations are equal to the product of the polarizations of the constituent subcircuits. These claims are equivalent whenever $n \gg 1$, because $\gamma(c) = S(c) + \mathcal{O}(\nicefrac{1}{4^n})$ and $\lambda(\mathcal{E}) = F(\mathcal{E}) + \mathcal{O}(\nicefrac{1}{4^n})$, \ie, Eqs.~(\ref{eq:gamma1})-(\ref{eq:gamma3}) imply the simpler results stated in the main text up to $\mathcal{O}(\nicefrac{1}{4^n})$ corrections.

The aim of the remainder of this section is to prove Eqs.~(\ref{eq:gamma1})-(\ref{eq:gamma3}), provide equations for $\Delta_1$ and $\Delta_2$, and argue why they are small. We can only prove these equations under some assumptions about the errors in our circuits, and so we will first state what those assumptions are.

\subsubsection{Assumptions}\label{sec:assumptions}
Proving Eqs.~(\ref{eq:gamma1})-(\ref{eq:gamma3}) rigorously requires a number of assumptions, which essentially formalize the notion of ``physically relevant situations''. We collect these assumptions here:
\begin{enumerate}
    \item We assume the standard ``Markovian'' model for errors in quantum computers \cite{Nielsen2020-lt,blumekohout_2021-tr}. Imperfect initialization into $\ket{0}\bra{0}^{\otimes n}$ is represented by an $n$-qubit state $\rho$. An imperfect measurement in the computational basis is represented by a positive-operator valued measure $\{E_x\}$, \ie, $E_x$ are positive operators satisfying $\sum_x E_x =\mathbb{I}$, where $E_x$ is a measurement effect corresponding to the result $x$. An imperfect implementation of a circuit $c=l_d\cdots l_2l_1$ is a completely positive and trace preserving superoperator $\phi(c)$. Moreover, $\phi(c) = \phi(l_d)\cdots \phi(l_2)\phi(l_1)$, where $\phi(l_i)$ are superoperators for the layers. The probability of observing $x$ when running circuit $c$ is then given by
    \begin{equation}
        P_x(c) = \textrm{Tr}\left\{ E_x \phi(c)[\rho]\right\}.
    \end{equation}
    While this assumption ignores non-Markovian and context-dependent errors, it has become standard in the literature due to the balance it strikes between accuracy and tractability. 
    
    \item All three circuit ensembles used in our procedure are defined with a layer of single qubit gates that randomize the initial state and measurement basis; $L$ and $L_{\textrm{rev}}$ in Eqs. (\ref{eq:M1})-(\ref{eq:M3}). We model all errors in state preparation and $L$ as an $L$-independent stochastic Pauli channel 
    $\mathcal{E}_{\rm sp}$ that occurs just after $L$, and all errors in the measurement and $L_{\textrm{rev}}$ as an $L$-independent stochastic Pauli channel that occurs just before $L_{\textrm{rev}}$. This is a stronger assumption that the general Markovian model assumption above, but this or similar assumptions are common in the literature. It is reasonable when single-qubit gate errors are small and bit flips dominate the state preparation and measurement errors.
    
    \item Randomized compilation requires factoring a circuit into alternating layers composed of single qubit gates and two-qubit gates; $\tilde{c} = l_{\tilde{d}}e_{\tilde{d}-1}l_{\tilde{d}-1}\cdots e_1l_1$, with $l_i$ the layers with single qubit gates, and $e_i$ the layers with two-qubit gates. As discussed in the main text, randomized compilation replaces each $l_i$ with a Pauli-randomized layer of single qubit gates. We model the errors associated to the single qubit layers as independent of layer, meaning that these errors can be entirely absorbed into the error channels on the two-qubit gates. This assumption is reasonable whenever two-qubit gate errors are much larger than single-qubit gate errors.  
\end{enumerate}

While these assumptions are necessary for our proof of \eqref{eq:chi2-F-theory} (by proving Eqs.~(\ref{eq:gamma1})-(\ref{eq:gamma3}) and providing formulae for $\Delta_1$ and $\Delta_2$), which relates $\chi_F(c)$ to $F(c)$, we do not believe they are all essential to this relationship. The numerical evidence presented in the main text suggests assumptions (2) and (3) are not essential, although we have not pursued this further in this work. Strict Markovianity (assumption 1) is also not necessary, although some forms of non-Markovianity will impact the reliability of MCFE (e.g., errors that get worse over the course of a circuit) and we do not attempt to create a theory here for the kinds of non-Markovianity that MCFE can tolerate.

In addition to the three assumptions above, our theory employs two (closely related) approximations, which are the cause of our $\Delta_1$ and $\Delta_2$ error terms. We now state these approximations.
\begin{enumerate}
    \item Consider an error channel composed of a product of unitarily rotated Pauli stochastic channels; \ie~ $\Lambda = \mathcal{U}_k \mathcal{S}_k \mathcal{U}_k\dg \cdots \mathcal{U}_1 \mathcal{S}_1 \mathcal{U}_1\dg$, where $\mathcal{S}_i$ are stochastic Pauli channels and $\mathcal{U}_i$ are unitaries. Our proof requires several properties related to the polarization of such a channel and its product with other channels. Specifically, we make use the following two approximations: 
    \begin{enumerate}
        \item $\lambda(\Lambda) \approx \prod_i \lambda(\mathcal{S}_i)$,
        \item $\lambda(\Lambda \mathcal{E}) \approx \lambda(\Lambda) \lambda(\mathcal{E})$, where $\mathcal{E}$ is an arbitrary error channel.
    \end{enumerate}
    Sec.~\ref{sec:lambda_comp} is dedicated to justifying these approximations.
    \item The final approximation we need relates the polarization of the error channel of a randomized compiled circuit to that of the error channel of the randomized compiled reverse circuit: $\lambda( \mathcal{E}(f_{\rm rc}(\tilde{c})) ) \approx \lambda( \mathcal{E}(f_{\rm rc}(\tilde{c}_{\rm rev})))$, where $\mathcal{E}(c)$ is the error channel for the circuit $c$. This approximation can be justified from our first approximation, but it is also separately justified in Sec. \ref{sec:M2_prop}. Our argument for why the error in this approximation is small follows from noting that the two-qubit gates in these randomly compiled circuits are self-inverse and Clifford.  
\end{enumerate}

\subsubsection{The state preparation and measurement ensemble}
First we analyze the randomized state preparation and measurement mirror circuit ensemble, $M_3(c)$. For these circuits
\begin{equation}
  P_x(M_3(c)) = \textrm{Tr}\left\{E_x \phi(f_{\textrm{rc}}(L_{\textrm{rev}}L))[\rho]\right\}.
\end{equation}
Letting $\mathcal{L}=\mathcal{U}(L)$, and by applying our assumptions (most notably, assumption 2) stated in Sec.~\ref{sec:assumptions}, we have that
\begin{align}
  P_x(M_3(c)) &= \bra{x} \mathcal{L}^{\dagger} \mathcal{E}_{\textrm{m}}\mathcal{E}_{\textrm{sp}}\mathcal{L}\mathcal{P}[\ket{0}\bra{0}]\ket{x} \\
&= \bra{x} \mathcal{L}^{\dagger} \mathcal{E}_{\textrm{m}}\mathcal{E}_{\textrm{sp}}\mathcal{L}[\ket{y}\bra{y}]\ket{x}, \label{eq:pxm3}
\end{align}
where $\mathcal{P}$ is the superoperator for a uniformly random Pauli operator, and $y$ is a uniformly random bit string. Therefore
\begin{align}
    S_3 
    &\equiv \mathbb{E}\{S(M_3(c))\},\\
    & = \mathbb{E}\left\{ \sum_{k=0}^{n}\left(-\frac{1}{2}\right)^{k} \sum_{h(x,y)=k} \bra{x} \mathcal{L}^{\dagger} \mathcal{E}_{\textrm{m}}\mathcal{E}_{\textrm{sp}}\mathcal{L}[\ket{y}\bra{y}]\ket{x} \right\}, \\
    & = \mathbb{E}_y\left\{ \sum_{x}\left(-\frac{1}{2}\right)^{h(x,y)} \mathbb{E}_{\mathcal{L}}(\bra{x} \mathcal{L}^{\dagger} \mathcal{E}_{\textrm{m}}\mathcal{E}_{\textrm{sp}}\mathcal{L}[\ket{y}\bra{y}]\ket{x}) \right\}, \\
    &= F(\mathcal{E}_{\textrm{m}}\mathcal{E}_{\textrm{sp}}). \label{eq:S3Fspam}
\end{align}
The second equality is obtained by substituting \eqref{eq:pxm3} into the definition of $S(\cdot)$ in \eqref{eq:S} [note that here $\sum_{h(x,y)=k}$ denotes summation over all bit strings $x$ and $y$ such that $h(x,y)=k$]. The third equality holds by the linearity of the expectation value, and the fourth equality holds from \eqref{eq:local-twirl-F} (which holds for any value of $y$, and so it holds when averaging over all values of $y$). By subtracting 1 from each side of $S_3 = F(\mathcal{E}_{\textrm{m}}\mathcal{E}_{\textrm{sp}})$ and then multiplying both sides by $\nicefrac{4^n}{4^n-1}$, using \eqref{eq:gamma} and \eqref{eq:polarization} we obtain
\begin{equation}
    \gamma_3 = \lambda(\mathcal{E}_{\textrm{m}}\mathcal{E}_{\textrm{sp}}).\label{eq:gamma3=gammaspam}
\end{equation}
This is \eqref{eq:gamma3} with 
\begin{equation}
\lambda_0 \equiv \lambda(\mathcal{E}_{\textrm{m}}\mathcal{E}_{\textrm{sp}}).\label{eq:gamma0}
\end{equation}

\subsubsection{Randomized compilation}
Deriving \eqref{eq:gamma1} and \eqref{eq:gamma2} relies on the theory of randomized compilation \cite{Wallman2016-rd}, and so we now take a brief diversion to review that theory in the context of our notation. Randomized compilation is a key ingredient in our method, because it converts general errors into stochastic Pauli channels. Consider any circuit $\tilde{c}$ of the form of \eqref{eq:alternating-circuit}, \ie, $\tilde{c} = l_{\tilde{d}}e_{\tilde{d}-1}l_{\tilde{d}-1}\cdots e_1l_1$. Because we assume layer-independent error channels for the single-qubit gate layers (the $l_i$), their errors can be absorbed into the errors on the two-qubit gate layers (the $e_i$) and we can therefore model $\phi(l_i)$ by $\mathcal{U}(l_i)$, with a corresponding redefinition of $\phi(e_i)$ that we do not explicitly denote. 
Now that we have argued that our assumptions enable us to let $\phi(l_i)=\mathcal{U}(l_i)$ for any single-qubit layer $l_i$, and defining $\mathcal{E}(e_i)$ by $\phi(e_i) = \mathcal{U}(e_i)\mathcal{E}(e_i)$, we have that
\begin{equation}
\phi(f_{\textrm{rc}}(\tilde{c})) = \cdots \mathcal{U}(e_2)\mathcal{E}(e_2)\mathcal{P}_{2}\mathcal{U}(l_2)\mathcal{P}^*_{1}\mathcal{U}(e_1)\mathcal{E}(e_1)\mathcal{P}_{1}\mathcal{U}(l_1),
\end{equation}
where the $\mathcal{P}_i$ are random Pauli operators and $\mathcal{P}_i^* = \mathcal{U}(e_i)\mathcal{P}_i\mathcal{U}(e_i)^{\dagger}$. Therefore
\begin{equation}
\phi(f_{\textrm{rc}}(\tilde{c})) = \cdots \mathcal{U}(e_2)\mathcal{P}_{2}\mathcal{E}(e_2)\mathcal{P}_{2}\mathcal{U}(l_2)\mathcal{U}(e_1) \mathcal{P}_{1}\mathcal{E}(e_1)\mathcal{P}_{1}\mathcal{U}(l_1),
\end{equation}
 
Averaging over the Pauli operators results in Pauli twirling of the error channels for the two-qubit gate layers. In particular, the mean of $\phi(f_{\textrm{rc}}(\tilde{c}))$ over all but the final random Pauli operator, \ie, the superoperator
\begin{equation}
 \bar{\phi}(f_{\textrm{rc}}(\tilde{c})) = \mathbb{E}_{\mathcal{P}_i,i<d}\{\phi(f_{\textrm{rc}}(\tilde{c}))\},
\end{equation}
is given by \cite{Wallman2016-rd}
\begin{equation}
 \bar{\phi}(f_{\textrm{rc}}(\tilde{c})) =\mathcal{P}_{\tilde{d}} \mathcal{U}(l_{\tilde{d}}) \prod_{k=1}^{\tilde{d}-1} \mathcal{U}(e_k)\mathcal{S}(e_k) \mathcal{U}(l_k),
\end{equation}
where $\mathcal{S}(e_k)$ is a stochastic Pauli channel given by
\begin{equation}
\mathcal{S}(e_k) = \mathbb{E}_{\mathcal{P}}\{ \mathcal{P} \mathcal{E}(e_k) \mathcal{P}\}. 
\end{equation}

\subsubsection{The $M_1(c)$ ensemble}
Turning to the $M_1(c)$ ensemble, we now we show that $\gamma_1 = \lambda_0 \lambda(\tilde{c}_{\textrm{rev}})\lambda(c) + \Delta_1$ where $\Delta_1$ is small, with 
\begin{equation}
   \tilde{c}_{\textrm{rev}} = l_{1,\textrm{rev}}e_1\dots l_{\tilde{d}-1,\textrm{rev}}e_{\tilde{d}-1} l_{\tilde{d},\textrm{rev}}. 
\end{equation}
Consider the superoperator $\phi(f_{\textrm{rc}}(\tilde{c}_{\textrm{rev}})c)$ which can be decomposed as 
\begin{equation}
 \phi(f_{\textrm{rc}}(\tilde{c}_{\textrm{rev}})c) = \phi(f_{\textrm{rc}}(\tilde{c}_{\textrm{rev}})) \mathcal{U}(c) \mathcal{E}(c).
 \end{equation}
We now apply the theory of randomized compilation, and our assumptions (Sec.~\ref{sec:assumptions}), to derive a formula for the action of $M_1(c)$ averaged over randomized compilations. The superoperator for $M_1(c)$ obtained when averaging over all Pauli operators in the randomized compilation of $\tilde{c}_{\textrm{rev}}$ except the last Pauli operator, is given by
\begin{align}
   \bar{\phi}(M_1(c)) &=  \mathcal{P}\mathcal{L}^{\dagger}\mathcal{E}_{\textrm{m}} 
   \bar{\mathcal{E}}(\tilde{c}_{\textrm{rev}})  \mathcal{U}^{\dagger}(c)\mathcal{U}(c) \mathcal{E}(c) \mathcal{E}_{\textrm{sp}}  \mathcal{L},
\end{align}
where $\mathcal{P}$ is a random Pauli, and 
\begin{equation}
    \bar{\mathcal{E}}(\tilde{c}_{\textrm{rev}}) = \mathcal{U}_1 \mathcal{S}_1 \mathcal{U}_1^{\dagger} \cdots \, \mathcal{U}_{\tilde{d}-1} \mathcal{S}_{\tilde{d}-1} \mathcal{U}_{\tilde{d}-1}^{\dagger}, \label{eq:barErev}
\end{equation}
with $\mathcal{S}_i=\mathcal{S}(e_{i})$, and $\mathcal{U}_i = \mathcal{U}(l_{1,\textrm{rev}}e_1l_{2,\textrm{rev}} \cdots e_i)$. By defining $\tilde{\mathcal{L}} =\mathcal{L}\mathcal{P}$ (which is identically distributed to $\mathcal{L}$), we then have that
\begin{align}
   \bar{\phi}(M_1(c)) &=  \tilde{\mathcal{L}}^{\dagger}\mathcal{E}_{\textrm{m}} 
   \bar{\mathcal{E}}(\tilde{c}_{\textrm{rev}})  \mathcal{E}(c) \mathcal{E}_{\textrm{sp}}  \tilde{\mathcal{L}}\mathcal{P}.
\end{align}
We therefore have that 
\begin{align}
   P_x[M_1(c)]& =  \bra{x} \tilde{\mathcal{L}}^{\dagger}\mathcal{E}_{\textrm{m}} 
   \bar{\mathcal{E}}(\tilde{c}_{\textrm{rev}})  \mathcal{E}(c)   \mathcal{E}_{\textrm{sp}}  \tilde{\mathcal{L}}\mathcal{P}[\ket{0}\bra{0}] \ket{x},\\
   &=  \bra{x} \tilde{\mathcal{L}}^{\dagger}\mathcal{E}_{\textrm{m}} 
   \bar{\mathcal{E}}(\tilde{c}_{\textrm{rev}})  \mathcal{E}(c)  \mathcal{E}_{\textrm{sp}}  \tilde{\mathcal{L}}[\ket{y}\bra{y}] \ket{x},
\end{align}
where $y$ is a uniformly random bit string. This equation has the same form as \eqref{eq:pxm3}, with $\mathcal{E}_{\textrm{m}}  \mathcal{E}_{\textrm{sp}}  \to \mathcal{E}_{\textrm{m}} 
   \bar{\mathcal{E}}(c_{\textrm{rev}})  \mathcal{E}(c)  \mathcal{E}_{\textrm{sp}} $. Therefore, we can apply Eqs.~(\ref{eq:pxm3})-(\ref{eq:gamma3=gammaspam}) to obtain
\begin{equation}
    \gamma_1 = \lambda\left( \mathcal{E}_{\textrm{m}} 
   \bar{\mathcal{E}}(\tilde{c}_{\textrm{rev}})  \mathcal{E}(c)   \mathcal{E}_{\textrm{sp}} \right).
\end{equation}
We now replace the right hand side of this equation---the polarization of a production of channels---with product of those channel's polarizations, with a correction factor $\Delta_1$ that encompasses the error in this approximation, i.e., we write
\begin{align}
\gamma_1 
   &=  \lambda\left( \mathcal{E}_{\textrm{m}}  \mathcal{E}_{\textrm{sp}}  \right)
   \lambda\left( \bar{\mathcal{E}}(\tilde{c}_{\textrm{rev}}) \right) \lambda\left(  \mathcal{E}(c) \right) + \Delta_1 , \label{eq:prod-pol-m3}
  \end{align}
  where $\Delta_1$ is simply
  \begin{equation}
      \Delta_1 \equiv   \lambda\left( \mathcal{E}_{\textrm{m}} 
   \bar{\mathcal{E}}(\tilde{c}_{\textrm{rev}})  \mathcal{E}(c)   \mathcal{E}_{\textrm{sp}} \right)  -  \lambda\left( \mathcal{E}_{\textrm{m}}  \mathcal{E}_{\textrm{sp}}  \right)
   \lambda\left( \bar{\mathcal{E}}(\tilde{c}_{\textrm{rev}}) \right) \lambda\left(  \mathcal{E}(c) \right).
  \end{equation}
We later argue that $\Delta_1$ is small. In summary, and as required, we have that 
\begin{align}
  \gamma_1 & = \lambda_0
\lambda_{\textrm{rev}} \lambda\left(c \right) + \Delta_1,
\end{align}
where we have used the definition for $\lambda_0$ in \eqref{eq:gamma0} and defined
\begin{equation}
  \lambda_{\textrm{rev}} = \lambda( \bar{\mathcal{E}}(\tilde{c}_{\textrm{rev}})) \label{eq:gammarev} .
\end{equation}
We have therefore derived \eqref{eq:gamma1}, as required. Before we explaining why $\Delta_1$ is (typically) small, we turn to $M_2(c)$.

\subsubsection{The $M_2(c)$ ensemble}
\label{sec:M2_prop}
We now consider the $M_2(c)$ circuit ensemble. We show that $\gamma_2 = \lambda_0 \lambda_{\textrm{rev}}^2 + \Delta_2$ where $\Delta_2$ is small. As with $M_1(c)$, we apply the theory of randomized compilation, and our assumptions (Sec.~\ref{sec:assumptions}). The theory or randomized compilation implies that the mean superoperator for $M_2(c)$, obtained by averaging over all Pauli operators in the randomized compilation of $\tilde{c}_{\textrm{rev}}$ and $\tilde{c}$ except the last Pauli operator, is given by
\begin{align}
   \bar{\phi}(M_2(c)) &=  \mathcal{P}\mathcal{L}^{\dagger}\mathcal{E}_{\textrm{m}} 
   \bar{\mathcal{E}}(\tilde{c}_{\textrm{rev}})  \mathcal{U}^{\dagger}(c)\mathcal{U}(c) \bar{\mathcal{E}}(\tilde{c}) \mathcal{E}_{\textrm{sp}} \mathcal{L},\\
   &= \tilde{\mathcal{L}}^{\dagger}\mathcal{E}_{\textrm{m}} 
   \bar{\mathcal{E}}(\tilde{c}_{\textrm{rev}})  \bar{\mathcal{E}}(\tilde{c})  \mathcal{E}_{\textrm{sp}} \tilde{\mathcal{L}} \mathcal{P},
    \label{eq:bar-phi-m2}
\end{align}
where $\bar{\mathcal{E}}(\tilde{c}_{\textrm{rev}})$ is defined in \eqref{eq:barErev}, and
\begin{equation}
    \bar{\mathcal{E}}(\tilde{c}) = \mathcal{V}_{\tilde{d}-1}^{\dagger}\mathcal{S}_{\tilde{d}-1} \mathcal{V}_{\tilde{d}-1}  \cdots \mathcal{V}_1^{\dagger} \mathcal{S}_1 \mathcal{V}_1,
\end{equation}
with $\mathcal{S}_i=\mathcal{S}(e_i)$ and $\mathcal{V}_i = \mathcal{U}(l_i \cdots e_1l_1)$.
We therefore have that 
\begin{align}
   P_x[M_2(c)]& =  \bra{x} \tilde{\mathcal{L}}^{\dagger}\mathcal{E}_{\textrm{m}} 
   \bar{\mathcal{E}}(\tilde{c}_{\textrm{rev}})  \bar{\mathcal{E}}(\tilde{c})  \mathcal{E}_{\textrm{sp}}\tilde{\mathcal{L}} \mathcal{P}[\ket{0}\bra{0}] \ket{x},\\
   &=  \bra{x} \tilde{\mathcal{L}}^{\dagger}\mathcal{E}_{\textrm{m}} 
   \bar{\mathcal{E}}(\tilde{c}_{\textrm{rev}})  \bar{\mathcal{E}}(\tilde{c})  \mathcal{E}_{\textrm{sp}} \tilde{\mathcal{L}} [\ket{y}\bra{y}] \ket{x},
\end{align}
where $y$ is a uniformly random bit string. This
equation has the same form as \eqref{eq:pxm3}, with $\mathcal{E}_{\textrm{m}}\mathcal{E}_{\textrm{sp}} \to \mathcal{E}_{\textrm{m}} 
   \bar{\mathcal{E}}(\tilde{c}_{\textrm{rev}})  \bar{\mathcal{E}}(\tilde{c}) \mathcal{E}_{\textrm{sp}}$. Therefore, we can apply Eqs.~(\ref{eq:pxm3})-(\ref{eq:gamma3=gammaspam}) to obtain
\begin{equation}
    \gamma_2 = \lambda\left( \mathcal{E}_{\textrm{m}} 
   \bar{\mathcal{E}}(\tilde{c}_{\textrm{rev}})  \bar{\mathcal{E}}(\tilde{c}) \mathcal{E}_{\textrm{sp}} \right).\label{eq:gamma2exact}
\end{equation}
By defining 
\begin{equation}
    \Delta_2 \equiv  \lambda\left( \mathcal{E}_{\textrm{m}} 
   \bar{\mathcal{E}}(\tilde{c}_{\textrm{rev}})  \bar{\mathcal{E}}(\tilde{c}) \mathcal{E}_{\textrm{sp}} \right) 
     - \lambda\left( \mathcal{E}_{\textrm{m}}\mathcal{E}_{\textrm{sp}} \right) \lambda\left( \bar{\mathcal{E}}(\tilde{c}_{\textrm{rev}}) \right)^2,
\end{equation}
which we will argue below is small, we can rewrite \eqref{eq:gamma2exact} as 
\begin{align}
   \gamma_2  
      & = \lambda\left( \mathcal{E}_{\textrm{m}}\mathcal{E}_{\textrm{sp}} \right)
   \lambda\left( \bar{\mathcal{E}}(\tilde{c}_{\textrm{rev}}) \right)^2 + \Delta_2, \label{eq:prod-pol-m2} \\
   & = \lambda_0
\lambda_{\textrm{rev}}^2 + \Delta_2.
\end{align}
Here we have used the definition for $\lambda_0$ in \eqref{eq:gamma0} and the definition for $\lambda_{\textrm{rev}}$ in \eqref{eq:gammarev}. This is \eqref{eq:gamma2}, as required.

The error term $\Delta_2$ can be thought of as coming from two approximations: first taking the approximation that
\begin{align}
    \lambda\left( \mathcal{E}_{\textrm{m}} 
   \bar{\mathcal{E}}(\tilde{c}_{\textrm{rev}})  \bar{\mathcal{E}}(\tilde{c})  \mathcal{E}_{\textrm{sp}} \right) 
   & \approx   \lambda\left( \mathcal{E}_{\textrm{m}} \mathcal{E}_{\textrm{sp}}  \right)
   \lambda\left( \bar{\mathcal{E}}(\tilde{c}_{\textrm{rev}}) \right) \lambda\left( \bar{\mathcal{E}}(\tilde{c}) \right), \label{eq:m2-polprod}
\end{align}
and then substituting in the approximation that 
\begin{equation}
\lambda( \bar{\mathcal{E}}(\tilde{c}_{\textrm{rev}}) ) \approx \lambda( \bar{\mathcal{E}}(\tilde{c})). \label{eq:approx:c:rev}
\end{equation}
Both approximations can be justified using the approximation that the polarization of a product of (rotated) stochastic Pauli channels is almost equal to the product of those channel's polarizations---which we address in detail below---but first we consider an alternative justification for the latter of these approximations [\eqref{eq:approx:c:rev}], as it is arguably particularly low-error. Using $\mathcal{U}_i^{\dagger} = e_i\mathcal{V}_i$, we have that
\begin{align}
    \mathcal{E}^{\dagger}(\tilde{c}_{\textrm{rev}}) &= \mathcal{U}_{\tilde{d}-1} \mathcal{S}_{\tilde{d}-1} \mathcal{U}_{\tilde{d}-1}^{\dagger} \cdots  \mathcal{U}_{1} \mathcal{S}_{1} \mathcal{U}_{1}^{\dagger} , \\
    &= \mathcal{V}_{\tilde{d}-1}^{\dagger} \mathcal{S}'_{\tilde{d}-1}  \mathcal{V}_{\tilde{d}-1}
    \cdots  \mathcal{V}_{1}^{\dagger} \mathcal{S}'_1 \mathcal{V}_{1} ,
\end{align}
where $\mathcal{S}'_i = \mathcal{U}(e_i)^{\dagger}\mathcal{S}_{i} \mathcal{U}(e_i)$. If $\mathcal{S}'_i = \mathcal{S}_i$ [which is true if $\mathcal{E}(e_i)$ commutes with $\mathcal{U}(e_i)$] then $\mathcal{E}^{\dagger}(\tilde{c}_{\textrm{rev}}) = \mathcal{E}(\tilde{c})$, and so \eqref{eq:approx:c:rev} holds exactly. More generally, conjugating the Pauli channel $\mathcal{S}_i$ by $\mathcal{U}(e_i)$ simply permutes the error rates of $\mathcal{S}_i$ ($e_i$ contains only Clifford gates). This can have some impact on the rate that stochastic errors on different layers cancel---impacting the polarization of the composite error map and meaning that $\lambda( \bar{\mathcal{E}}(\tilde{c}_{\textrm{rev}}) )$ does not exactly equal $\lambda( \bar{\mathcal{E}}(\tilde{c}))$---but typically this impact will be very small (see theory below). 

\subsubsection{Approximate formula for the polarization of composed channels}
\label{sec:lambda_comp}
Our theory supporting the claim that $\chi_F(c) \approx F(c)$ relies on the claim that the polarization of the product of certain sequences of error channels is approximately equal to the product of those channel's polarizations, i.e., this is the reason we claim that $\Delta_1$ and $\Delta_2$ in the above theory are very small. 
We now justify this approximation, enabling us to argue that $\Delta_1$ and $\Delta_2$  are very small.
Consider a composite error channel
\begin{equation}
\Lambda = \mathcal{U}_k\mathcal{S}_k \mathcal{U}_k^{\dagger} \cdots \mathcal{U}_1\mathcal{S}_1 \mathcal{U}_1^{\dagger},
\end{equation}
where the $\mathcal{U}_i$ are unitary superoperators and the $\mathcal{S}_i$ are stochastic Pauli channels. Our theory relies on the approximation:
\begin{equation}
    \lambda(\Lambda) \approx \prod_i \lambda(\mathcal{S}_i),\label{eq:pol-approx}
\end{equation} 
where $\lambda$ is the polarization defined in \eqref{eq:polarization}, and errors in this approximation cause $\Delta_1$ and $\Delta_2$ to deviate from zero. Note that this equation holds exactly when every $\mathcal{S}_i$ is an $n$-qubit depolarizing channel, and that $\lambda(\mathcal{A}_1\cdots \mathcal{A}_k)$ is not well approximated by $\lambda(\mathcal{A}_1)\cdots \lambda(\mathcal{A}_k)$ for general error channels $\mathcal{A}_i$. We now upper- and lower-bound 
\begin{equation}
 \Delta  = \lambda(\Lambda) - \prod_i \lambda(\mathcal{S}_i), 
\end{equation}
and we argue that $|\Delta|$ is small in most physically relevant situations and the context of MCFE, i.e., we conjecture it is typically much smaller than our upper bound on $|\Delta|$.

The action of a sequence of $k$ unitarily rotated stochastic Pauli channels can be unravelled, \ie, $\Lambda$ can be written as a weighted sum over $k4^n$ terms consisting of all possible combinatons of errors occuring in the $k$ channels. Specifically, 
\begin{equation}
\Lambda=\sum_{\mathcal{P}_1\cdots\mathcal{P}_k} (r_{k,\mathcal{P}_k}\cdots r_{1,\mathcal{P}_1})\mathcal{U}_k\mathcal{P}_k\mathcal{U}_k^{\dagger} \cdots \mathcal{U}_1\mathcal{P}_1\mathcal{U}_1^{\dagger}
\end{equation}
where $r_{j,\mathcal{P}}$ is the probability that $\mathcal{S}_j$ applies the Pauli $\mathcal{P}$, and so $r_{j,\mathbb{I}}$ is $\mathcal{S}_j$'s fidelity where $\mathbb{I}$ is the identity Pauli. Using this expansion, and the fact that $F(\mathcal{U}\mathcal{P}\mathcal{U}^{\dagger})=0$ for any non-identity Pauli, we therefore find that 
\begin{equation}
F(\Lambda) = \prod_iF(\mathcal{S}_i) + \mathcal{O}(k^2\epsilon^2)
\end{equation}
where $1-F(\mathcal{S}_i)=\mathcal{O}(\epsilon)$ for all $i$. Therefore
\begin{equation}
|\Delta| \leq \mathcal{O}(k^2\epsilon^2) \label{eq:Delta1storder}
\end{equation}
when $ F(\mathcal{S}_i) = 1 - \mathcal{O}(\epsilon)$, \ie, when the infidelities of the stochastic Pauli channels $\mathcal{S}_i$ are all order $\epsilon$. Therefore,  $|\Delta|$ is negligible when $k\epsilon \ll 1$. This is sufficient to show that $\Delta_1, \Delta_2 \ll 1$, and that $\Delta_1, \Delta_2 \ll 1 - F(c)$, in the regime of $F(c)\approx 1$. 

We now address the case when $k\epsilon$ is not small---which will typically be the case when $F(c)$ is not close to one---and so the bound of \eqref{eq:Delta1storder} is not small. First, we lower bound $\Delta$. 
For a fixed value of the $F(\mathcal{S}_i)$, the minimal value for $\lambda(\Lambda)$ is obtained when error cancellation is minimized, \ie, when the probability that two or more errors that occur in the sequence $\Lambda$ compose to the identity is minimized. The probability of an error induced by $\mathcal{S}_i$ being cancelled by an error in $\mathcal{S}_j$ can be zero, \eg, there is no probability of error cancelation in the channel $\mathcal{S}_1\mathcal{S}_2$ if the set of errors with non-zero probability for $\mathcal{S}_1$ and $\mathcal{S}_2$ have no overlap. In this case $F(\Lambda)= \prod_i F(S_i)$, as the fidelity of a Pauli channel is equal to the probability that it applies the identity Pauli. As $F(\mathcal{E}) = \lambda(\mathcal{E}) + \mathcal{O}(\nicefrac{1}{4^n})$ we therefore find that $\Delta \geq -\delta_{\textrm{lower}}$ where $\delta_{\textrm{lower}} = \mathcal{O}(\nicefrac{1}{4^n})$.

Next we upper bound $\Delta$. In general $\Delta \leq \nicefrac{1}{2}$. This bound is saturated in the large $k$ limit with $\mathcal{U}_i=\mathbb{I}$ and with all $\mathcal{S}_i$ equal to a stochastic Pauli channel $\mathcal{S}$ with a distribution over Pauli errors that has support only on one non-identity Pauli. That is, $\mathcal{S}$ applies the identity with probability $F(\mathcal{S})$ and otherwise it applies some fixed Pauli $\mathcal{P}$. This maximizes $\Delta$ as $\mathcal{S}$ has a maximal error cancellation rate, and $\mathcal{S}^k$ converges to a uniform distribution over $\mathbb{I}$ and $\mathcal{P}$---so $\lambda(S^k)=\nicefrac{1}{2}$ as $k \to \infty$. 

This example uses a physically implausible error channel with a maximally sparse error probability vector. We conjecture that $\Delta \ll \nicefrac{1}{2}$ whenever each $\mathcal{S}_i$'s error probability vector $\vec{r}$ is not close to maximally sparse.  We now provide some support for this conjecture, and which we note can be proven to be true when the $\mathcal{U}_i$ are all Clifford operators (which will not typically be true in MCFE).
To do so, we will use an alternative formula for $\lambda(\mathcal{E})$ that holds for all completely positive and trace preserving $\mathcal{E}$:
\begin{equation}
    \lambda(\mathcal{E}) = \frac{\textrm{Tr}\, [\mathcal{E}]_{\textrm{u}}}{4^n-1} , \label{eq:pol-trace}
\end{equation}
where $[\mathcal{E}]_{\textrm{u}}$ is the unital component of $\mathcal{E}$.
We assume that $F(\mathcal{S}_i) \geq 0.5$ for all $i$, as then each $\mathcal{S}_i$ is a positive semi-definite matrix, and hence each $\mathcal{U}_i \mathcal{S}_i \mathcal{U}_i^{\dagger}$ is a positive semi-definite matrix. This allows us to use the following result. Any positive semi-definite matrices $A_i$, with $i=1,2,\dots,k$, satisfy \cite{Shebrawi2013-hh}
\begin{equation}
\textrm{Tr}\left(A_{1}A_{2}\cdots A_k \right) \leq \prod_{i=1}^{k} \textrm{Tr}(A_i^k)^{\frac{1}{k}}.
\end{equation}
By using this bound and \eqref{eq:pol-trace}, we obtain
\begin{equation}
\lambda(\mathcal{U}_k\mathcal{S}_k \mathcal{U}_k^{\dagger} \cdots \mathcal{U}_1\mathcal{S}_1 \mathcal{U}_1^{\dagger}) \leq \prod_{i=1}^{k} \lambda(\mathcal{S}_i^k)^{\frac{1}{k}} \leq \lambda(\mathcal{S}^k),
\end{equation}
where $\mathcal{S}$ is the $\mathcal{S}_i$ for which $\lambda(S_i^k)$ is maximized.
Note that both inqualities are saturated when $\mathcal{U}_i =\mathbb{I}$ for all $i$ and $\mathcal{S} = \mathcal{S}_i$ for all $i$ and some $\mathcal{S}$. Therefore, if $\mathcal{S}_i=\mathcal{S}$ for some $\mathcal{S}$ and all $i$, we have that
\begin{equation}
    \Delta \leq \lambda(S^k) - \lambda(S)^k.
\end{equation}
The difference $\lambda(S^k) - \lambda(S)^k$ can be bounded by the sparsity of $\mathcal{S}$'s vector of Pauli error rates $\vec{r}$, decreasing to zero as the sparsity decreases \cite{Proctor2021-bq}. Note high sparsity is physically unlikely, \eg, a tensor product of single-qubit depolarizing channels with similar error rates has low sparsity.

The final result that our theory relies on is the following:
\begin{equation}
    \label{eq:lambdalambdalambda}
    \lambda(\Lambda \mathcal{E}) \approx \lambda(\Lambda)\lambda(\mathcal{E}),
\end{equation}
where $\Lambda$ is as above, and $\mathcal{E}$ is any error channel [we used this in deriving \eqref{eq:prod-pol-m3}, as the error channel for $c$ is arbitrary, and so the errors in this approximation contribute to $\Delta_1$]. To derive this, we consider the case of a single unitarily rotated stochastic Pauli channel, $\Lambda = \mathcal{U} \mathcal{S} \mathcal{U}^{\dagger}$. If \eqref{eq:lambdalambdalambda} approximately holds for such $\Lambda$ then the following argument can be iteratively applied to show that \eqref{eq:lambdalambdalambda} approximately holds for $\Lambda$ an arbitrary sequence of unitarily rotated Pauli channels. By using the cyclic property and linearity of the trace, we have that
\begin{align}
    \lambda(\mathcal{U} \mathcal{S} \mathcal{U}^{\dagger} \mathcal{E}) &= \frac{\textrm{Tr}\, [\mathcal{U} \mathcal{S} \mathcal{U}^{\dagger} \mathcal{E}]_{\textrm{u}}}{4^n-1} ,\\
     &= \mathbb{E}_{\mathcal{P}}
     \frac{\textrm{Tr}\, [\mathcal{U}\mathcal{P}\mathcal{U}^{\dagger}\mathcal{U} \mathcal{S} \mathcal{U}^{\dagger} \mathcal{E}\mathcal{U}\mathcal{P}\mathcal{U}^{\dagger}]_{\textrm{u}}}{4^n-1} ,\\
          &= \mathbb{E}_{\mathcal{P}}
     \frac{\textrm{Tr}\, [\mathcal{U}\mathcal{S}\mathcal{U}^{\dagger}\mathcal{U} \mathcal{P} \mathcal{U}^{\dagger} \mathcal{E}\mathcal{U}\mathcal{P}\mathcal{U}^{\dagger}]_{\textrm{u}}}{4^n-1} ,\\
               &= 
     \frac{\textrm{Tr}\, [\mathcal{U}\mathcal{S}\mathcal{U}^{\dagger}\mathcal{S}_{\mathcal{E}}]_{\textrm{u}}}{4^n-1} ,\\
    & = \lambda(\mathcal{U} \mathcal{S} \mathcal{U}^{\dagger} \mathcal{S}_{\mathcal{E}})\\
    & \approx \lambda(\mathcal{U} \mathcal{S} \mathcal{U}^{\dagger})\lambda( \mathcal{S}_{\mathcal{E}}) \label{eq:lambdaE6}\\
    & = \lambda(\mathcal{U} \mathcal{S} \mathcal{U}^{\dagger})\lambda(\mathcal{E}),
\end{align}
where $\mathcal{P}$ is a uniformly random Pauli superoperator,
\begin{equation}
\mathcal{S}_{\mathcal{E}} = \mathbb{E}_{\mathcal{P}}\{\mathcal{U} \mathcal{P} \mathcal{U}^{\dagger} \mathcal{E}\mathcal{U}\mathcal{P}\mathcal{U}^{\dagger}\}
\end{equation}
is a unitary rotation of a stochastic Pauli channel that is the Pauli twirl of a unitary rotation of $\mathcal{E}$---and so it satisfies $F(\mathcal{S}_{\mathcal{E}})=F(\mathcal{E})$---and the approximate equality of \eqref{eq:lambdaE6} holds from our above theory for the polarization of sequences of unitarily rotated stochastic Pauli channels.

\subsection{Simulations}\label{app:simulations}
In this section we provide further details for our simulations, the results of which are presented in Fig.~\ref{fig:simulations} and~\ref{fig:simulations-2}.

\subsubsection{The QAOA circuits}
We applied our technique to QAOA circuits using simulated data. A high-level description of $n$-qubit QAOA circuits is shown in Fig.~\ref{fig:simulations}A. The QAOA circuits we use are parameterized by an $n$-node problem graph, the number of algorithmic layers $p$, and two length $p$ vectors of angles $\vec{\alpha}, \vec{\beta} \in (-\pi,\pi]^{p}$. For the simulations presented in Fig.~\ref{fig:simulations}, we systemically vary $n$ and $p$ (setting $n=3,4,\dots,8$ and $p=1,2,5,8, 10$) and for each $(n,p)$ the remaining parameters for 10 different $n$-qubit, $p$-layer QAOA circuits are chosen by independently applying the following procedure. First we sample a random $n$-node (Erd\H{o}s-R\'enyi) graph with an edge inclusion probability of $p_{\textrm{edge}}=0.5$ and with a random weight $w_{ij}$ sampled independently and uniformly from $[0,1]$ assigned to each edge $(i,j)$. Then we sample each $\alpha_i$ and $\beta_i$ independently and uniformly at random from $(-\pi,\pi]$. For the simulations in Fig.~\ref{fig:simulations-2} we fix $p=1$ and $n=100$ and we sample 75 random $100$-node (Erd\H{o}s-R\'enyi) graph with an edge inclusion probability of $p_{\textrm{edge}}=0.1$
and without weights (\ie, we set $w_{ij}=1$ for all edges $(i,j)$). For each graph, we then sample $\alpha_1$ and $\beta_1$ independently and uniformly at random from $\{-\nicefrac{\pi}{2},0,\nicefrac{\pi}{2},\pi\}$. This latter procedure guarantees that the QAOA circuit can be implemented with only Clifford gates, which we require to implement the simulations for Fig.~\ref{fig:simulations-2} efficiently.

For a given $n$-node weighted graph, and a given $\vec{\alpha}$ and $\vec{\beta}$, a QAOA circuit is constructed as follows. The QAOA circuits consist of alternating applications of unitaries generated by two Hamiltonians $H_C$ and $H_D$. $H_C$ encodes the combinatorial optimization problem. $H_D$ is a driver Hamiltonian that does not commute with $H_C$. We choose the standard form for the driver Hamiltonian: $H_D=\sum_i X_i$. We select $H_C$ to encode the MaxCut problem on the given $n$-node weighted graph. This Hamiltonian $H_C$ consists of Ising terms $w_{ij}Z_iZ_j$ between all edges $(i,j)$ in the graph, where $w_{ij}$ is the problem graph's weight for edge $(i,j)$. The QAOA circuits consist of applying the unitary $e^{-i \alpha_i H_C}$ and then the unitary $e^{-i \beta_i H_D}$ for $i=1,2,\dots,p$. For any value of $\beta_i$,  $e^{-i \beta_i H_D}$ is a tensor product of single-qubit unitaries. Therefore each $H_D$ layer is implemented by a single layer of one-qubit gates. Each $e^{-i \alpha_i H_C}$ layer is implemented by compiling it into maximally parallelized $e^{-i \alpha_i w_{jk}Z_jZ_k}$ two-qubit unitaries, each of which is compiled into CNOT gates and single-qubit gates. Each QAOA circuit therefore consists of arbitrary single-qubit gates and CNOT gates.

\subsubsection{The simulated error models}\label{app:simulations-error-models}
For the results presented in Fig.~\ref{fig:simulations}, we simulated our MCFE procedure for each of the randomly sampled 1-8 qubit QAOA circuits (see above). The MCFE procedure is simulated for each of these QAOA circuits under four error models, which are sampled from four different error model families. These four families are denoted in the main text by ``S'', ``H'', ``H+S'' and ``H-2Q''. All four error models are defined using the error generator formalism of Ref.~\cite{blumekohout_2021-tr}, \ie, all rates specified are rates of elementary error generators in a post-gate error map. Each of our single-qubit gates is implemented using the sequence $Z(\theta)X_{\nicefrac{\pi}{2}}Z(\phi)X_{\nicefrac{\pi}{2}} Z(\psi)$ and we include errors only on the $X_{\nicefrac{\pi}{2}}$ gates. We include over-rotation Hamiltonian errors and Pauli stochastic errors on each $X_{\nicefrac{\pi}{2}}$ with qubit-dependent error rates sampled uniformly in the range $[0,\theta_1]$ and $[0,\epsilon_1]$, respectively. The stochastic error is then split over the three Pauli errors uniformly at random.  We include over-rotation Hamiltonian errors and Pauli stochastic errors on each CNOT gate with qubit-pair-dependent error rates sampled uniformly in the range $[0,\theta_2]$ and $[0,\epsilon_2]$, respectively. The stochastic error is then split over the 15 two-qubit Pauli errors at random. We include readout error by preceding a perfect measurement with a tensor product of single-qubit depolarizing channels with errors rates sampled independently and uniformly from the interval $[0,10^{-2}]$. In the ``H'' model family $\epsilon_1=\epsilon_2=0$, $\theta_1=1.25\times 10^{-1}$ and $\theta_2=2.5\times 10^{-1}$. In the ``S'' model family $\epsilon_1=1\times 10^{-2}$, $\epsilon_2=2\times 10^{-2}$ and $\theta_1=\theta_2=0$. In the ``H+S'' model family $\epsilon_1=5 \times 10^{-3}$, $\epsilon_2=10^{-2}$, $\theta_1=7.5 \times 10^{-2}$ and $\theta_2=1.25 \times  10^{-1}$. In the ``H-2Q'' model family $\epsilon_1=\epsilon_2=\theta_1=0$ and $\theta_2=2.5 \times 10^{-1}$. For each of 1-8 qubit QAOA circuits that we sampled (300 circuits in total), we constructed a unique model from each model family by sampling each of its error rates. All simulations were implemented using \texttt{pyGSTi} \cite{Nielsen2020-rd}, using a density matrix simulation. This is a strong simulation, \ie, it computes a circuit's output probability distribution (under the error model) rather than sampling from it. As our aim in these simulations is to compare $\chi_F(c)$ and $F(c)$ we do not add finite sample error by drawing $K$ samples from these distributions (so this is equivalent to $K \to \infty$).

For the results presented in Fig.~\ref{fig:simulations-2}, we simulated our MCFE procedure for each of the 75 randomly sampled 100-qubit QAOA circuits (see above) each with a different error model sampled from a family of Pauli stochastic error models. The circuit sampling is designed so that all of these circuit contain only Clifford gates. This makes it possible to efficiently simulate these circuits under stochastic Pauli error models, using a stochastic unravelling. We do so using a \texttt{pyGSTi} interface to the \texttt{CHP} Clifford circuit simulator \cite{Aaronson2004-ab}. A stochastic unravelling is a weak simulation, \ie, it generates samples from the circuit's output probability distribution rather than computing the distribution. We generate $K = 180$ samples from each simulated circuit's distribution. Our error model family consisted of Pauli stochastic errors on each $X_{\nicefrac{\pi}{2}}$ gate, each CNOT gate, and each qubit's readout, with error probabilities sampled uniformly from $[0,\epsilon_1']$ and $[0,\epsilon_2']$ and $[0,10^{-3}]$ respectively. To obtain a broad range of values for $F(c)$, we varied $\epsilon_1'$ and $\epsilon_2'$ from $\epsilon_1'=5\times 10^{-5}$ and $\epsilon_2'=1\times 10^{-4}$ up to $\epsilon_1'=2\times 10^{-3}$ and $\epsilon_2'=7\times 10^{-3}$. Each of the 75 QAOA circuits is simulated under a distinct error model sampled from this error family for one value of these sampling parameters, within the stated range. Note that the error models for  Fig.~\ref{fig:simulations} use a different error model parameterization to those of Fig.~\ref{fig:simulations-2}. In the former case stochastic errors are parameterized using the rates of the stochastic error generators in the error generator framework, and in the latter case they are parameterized by the probabilities of applying each possible Pauli error.

In Figs.~\ref{fig:simulations} and~\ref{fig:simulations-2} we compare $\chi_F(c)$, which is the process fidelity $F(c)$ predicted by our method, to the true value of $F(c)$. We calculate this true value of $F(c)$ using a sampling procedure. This procedure consists of simulating randomly sampled $M'(c) = L_{\textrm{rev}} c_{\textrm{rev}} c L$ circuits, and calculating their $\bar{S}$, with all parts of the circuit implemented perfectly except $c$ (which is simulated under the error model in question). In this paper, we prove that $\bar{S} = F(c)$ under these conditions. For the results of Fig.~\ref{fig:simulations},
we randomly sampled and simulated 2000 $M'(c)$ circuits to calculate each $F(c)$, with no finite sampling error added to each circuit's output distribution (equivalent to $K \to \infty$). For the results of Fig.~\ref{fig:simulations},
we randomly sampled and simulated 400 $M'(c)$ circuits to calculate each $F(c)$ with $K=180$ for each circuit. 

\end{document}